\documentclass[%
superscriptaddress,
pof,
 amsmath,
 amssymb,
preprint,%
 longbibliography,  
]{revtex4-2}

\usepackage{graphicx}
\usepackage{dcolumn}
\usepackage{bm}
\usepackage[mathlines]{lineno}

\usepackage[utf8]{inputenc}
\usepackage[T1]{fontenc}
\usepackage{mathptmx}
\usepackage{etoolbox}

\usepackage{enumerate}  
\usepackage{enumitem}   
\usepackage{tikz}       
\usepackage[linesnumbered,ruled,vlined]{algorithm2e} 
\usepackage{hyperref}   
\setlist{nosep}         
\usepackage{makecell}   
\usepackage{booktabs} 
\usepackage{multirow} 

\makeatletter
\def\@email#1#2{%
 \endgroup
 \patchcmd{\titleblock@produce}
  {\frontmatter@RRAPformat}
  {\frontmatter@RRAPformat{\produce@RRAP{*#1\href{mailto:#2}{#2}}}\frontmatter@RRAPformat}
  {}{}
}%
\makeatother
\begin{document}

\preprint{AIP/123-QED}

\title[Sample title]{Flux-controlled wall model for large eddy simulation integrating the compressible law of the wall}

\author{Youjie Xu}
\thanks{youjie.xu@tum.de}
\affiliation{Chair of Aerodynamics and Fluid Mechanics, TUM School of Engineering and Design, Technical University of Munich, Boltzmannstraße 15, 85748 Garching, Germany}

\author{Steffen J. Schmidt}
\affiliation{Chair of Aerodynamics and Fluid Mechanics, TUM School of Engineering and Design, Technical University of Munich, Boltzmannstraße 15, 85748 Garching, Germany}

\author{Nikolaus A. Adams}
\affiliation{Chair of Aerodynamics and Fluid Mechanics, TUM School of Engineering and Design, Technical University of Munich, Boltzmannstraße 15, 85748 Garching, Germany}
\affiliation{Munich Institute of Integrated Materials, Energy and Process Engineering, Technical University of Munich, Lichtenbergstraße 4a, 85748 Garching, Germany}

\date{\today}

\begin{abstract}
Recent advances in velocity and temperature transformations have enabled recovery of the law of the wall in compressible wall-bounded turbulent flows. Building on this foundation, a flux-controlled wall model (FCWM) for Large Eddy Simulation (LES) is proposed. Unlike conventional wall-stress models that solve the turbulent boundary layer equations, FCWM formulates the near-wall modeling as a control problem applied directly to the outer LES solution. It consists of three components: (1) the compressible law of the wall, (2) a feedback flux-control strategy, and (3) a shifted boundary condition. The model adjusts the wall shear stress and heat flux based on discrepancies between the computed and target transformed velocity and temperature, respectively, at the matching location. The proposed wall model is evaluated using LES of turbulent channel flows across a broad range of conditions, including quasi-incompressible cases with bulk Mach number \(M_b = 0.1\) and friction Reynolds number \(Re_\tau = 180 \sim 10{,}000\), and compressible cases with \(M_b = 0.74 \sim 4.0\) and bulk Reynolds number \(Re_b = 7667 \sim 34{,}000\). The wall-modelled LES reproduce mean velocity and temperature profiles in agreement with direct numerical simulation data. For all tested cases with \(M_b \leq 3\), the wall model achieves relative errors of \(|\epsilon_{C_f}| < 4.1\%\), \(|\epsilon_{B_q}| < 2.7\%\), and \(|\epsilon_{T_c}| < 2.7\%\) in friction coefficient, non-dimensional heat flux, and centerline temperature, respectively. In the quasi-incompressible regime, the wall model achieves \(|\epsilon_{C_f}| < 1\%\). Compared to the conventional equilibrium wall model, the proposed FCWM achieves higher accuracy in compressible turbulent channel flows without solving the boundary layer equations, thereby reducing computational cost.
\end{abstract}

\maketitle
\section{\label{section:introduction}Introduction}
Wall-bounded turbulent flows are common in applications such as wind farms \citep{Sørensen2011}, aircraft aerodynamics \citep{Slotnick2014,Goc2021}, and atmospheric flows \citep{Stoll2020}. These flows are typically characterized by high Reynolds numbers and multiscale turbulence \citep{Smits2011}. Compared to Direct Numerical Simulation (DNS) and Reynolds-Averaged Navier-Stokes (RANS) approach, LES achieves a balance between accuracy and computational cost by resolving large, energy-containing scales and modeling smaller, isotropic scales with a subgrid-scale (SGS) model.

According to \citet{Pope2000}, a reliable LES should resolve at least \(80\%\) of the turbulent kinetic energy (TKE). In wall-bounded turbulent flows, the size of energetic and dynamically important eddies decreases progressively toward the wall, particularly at high Reynolds numbers. Resolving these near-wall scales in LES requires grid resolution and time step size comparable to DNS \citep{Cabot2000}, which severely limits the application of LES to high-Reynolds-number flows in engineering. Many studies have examined the grid requirement for turbulence simulation \citep{Chapman1979, Choi2012, Rezaeiravesh2016, Yang2021}. The recent estimation by \citet{Yang2021} indicates that the computational cost of wall-resolved LES (WRLES) and DNS scale as \(Re^{1.86}\) and \(Re^{2.05}\) for a flat-plate boundary layer, respectively. In fact, the high cost of WRLES stems from resolving the inner layer, which accounts for only \(10\%\) of the boundary layer but consumes \(99\%\) of the grid points at \(Re = \mathcal{O} (10^6)\) \citep{Piomelli-Balaras}. To overcome this limitation, wall-modeled LES (WMLES) is employed, which resolves only the energy-containing scales in the outer layer on a coarse grid, while the dynamically important near-wall scales are fully modeled, with their effects on the outer flow imposed through approximate boundary conditions. The computational cost of WMLES scales with \(Re\) \citep{Yang2021}, making it an efficient choice for high-Reynolds-number flows.

Numerous WMLES approaches have been developed over the years. They are commonly categorized into hybrid LES/RANS methods and wall-stress models, depending on how the resolved and modeled regions are coupled \citep{Larsson2016}. In hybrid LES/RANS approach, LES is applied above an interface, while RANS is used below it \citep{Piomelli2003, Heinz2020}. In wall-stress models, LES extends all the way to the wall, with the wall model supplying the instantaneous shear stress and heat flux at the wall. Apart from this two categories, there are also a few other wall models, including the integral wall model \citep{Yang2015, Catchirayer2018}, the slip wall model \citep{Bose2014, Bae2019, Shi2022}, the control-based wall model \citep{Nicoud2001, Templeton2006, Templeton2008}, stochastic forcing \citep{Keating2006,Ying2025}, and those using machine learning approaches \citep{Yang2019, Bae2022, TabeJamaat2023, Lee2023, TabeJamaat2024, Lyu2024}. The reader is directed to reviews \citep{Cabot2000, Piomelli-Balaras, Piomelli2008, Larsson2016, Bose2018} for more comprehensive overview. Among these approaches, the wall-stress model and control-based wall model are directly related to the focus of the present study.

Wall-stress modeling in LES can be implemented using either turbulent boundary layer equations (TBLEs) or the law of the wall. In TBLE-based approach, the wall shear stress and heat flux are obtained by numerically solving the TBLEs \citep{Wang2002, Kawai2013}. When the unsteady and convective terms are assumed to be approximately in balance with pressure gradient, the TBLEs reduce to Ordinary Differential Equations (ODEs), forming the commonly used equilibrium-wall-model (EWM) \citep{Iyer2019, Kawai2012, Griffin2020}. Alternatively, integrating the momentum ODE across the logarithmic layer directly yields the log-law \citep{Iyer2019}. Thus, the wall model can also be applied by algebraically solving the log-law with the Newton-Raphson method \citep{Bocquet2012} or through a tabular approach \citep{Maheu2012}. Both methods fall under the category of algebraic or analytical wall models, which have been employed for near-wall modeling since the 1970s \citep{Deardorff1970, Schumann1975, Groetzbach1987, Piomelli1989} and have seen continued development in recent years \citep{Cai2021,Nuca2025}.

Nevertheless, the accuracy of WMLES depends not only on the wall model, but also on the numerical scheme and SGS model employed \citep{Cabot2000}. The coarse grid resolution inherent in WMLES inevitably introduces numerical and modeling errors across the first few off-wall cells, which are considered a primary source of the well-known log-layer mismatch (LLM) \citep{Kawai2012, Yang2017, Maejima2024}. To overcome this limitation, the control-based wall model proposed by \citet{Nicoud2001} formulates the near-wall modeling as a control problem. It accounts for the numerical and modeling errors by enforcing a physically significant log-law, thus removing the LLM. However, the computational cost of the original control-based wall model is relatively expensive, even with efficiency improvements \citep{Templeton2006, Templeton2008}.

In addition, the wall models introduced above are typically implemented for incompressible flows. Their application to compressible flows presents additional challenges, primarily due to the coupling between the momentum and energy equations and to viscous heating effects. \citet{Griffin2023} pointed out that, iteratively solving the coupled ODEs introduces higher degree of nonlinearity and can be difficult to converge in flows with steep temperature profile. In addition, the wall model accuracy also degrades in flows with strong heat transfer, as demonstrated by the EWM results \citep{Chen2022b, Griffin2023}. These challenges have motivated alterative approaches based on compressible transformations.

In recent years, one of the significant advances in the study of compressible wall-bounded turbulent flows has been the development of compressible law of the wall, including various velocity transformations \citep{Zhang2012, Trettel2016, Patel2016, Wu2017, Volpiani2020, Griffin2021, Hasan2023, Zhu2024, Xu2025a} and temperature transformations \citep{Patel2017, Chen2022a, Huang2023, Cheng2024a, Modesti2024, Xu2025b}. These transformations are designed to map the compressible velocity and temperature profiles to their incompressible counterparts, allowing existing incompressible modeling techniques to be extended to compressible flows. Additionally, motivated by the Strong Reynolds Analogy (SRA) \citep{Morkovin1962}, many temperature-velocity (TV) relations have been established since last century \citep{Walz1962, Duan2011, Zhang2014, Cheng2024b, Zhu2025a}, making it possible to obtain the mean temperature profile from the mean velocity distribution. The reader can refer to the recent review by \citet{Cheng2024c} for comprehensive discussion of near-wall modeling in compressible wall-bounded turbulent flows. These advancements have led to the development of many new wall models for simulating high-speed flows. Among these, at least three strategies have been explored. In the first strategy, the incompressible eddy viscosity model is augmented with the velocity transformation kernel to provide the compressible eddy viscosity in the ODE-based wall model \citep{Hendrickson2022, Hendrickson2023}. Similar approach is applied in the \(k-\omega\) Shear Stress Transport (SST) model by \citet{Hasan2025}. In the second strategy, the momentum ODE for incompressible flows is invoked, followed by an inverse velocity transformation and an algebraic TV-relation to obtain the compressible velocity, temperature, density, and viscosity profiles without solving the energy equation. The wall model by \citet{Griffin2023} follows this approach and demonstrates improved performance over the traditional ODE-based wall model in strong heat transfer scenarios. Furthermore, \citet{Chen2025} propose to inversely solve the temperature transformation by \citet{Cheng2024a}, thereby removing the dependence of TV-relation on boundary layer edge quantities. In the third strategy, the ODEs are completely avoided by inversely applying the velocity and temperature transformations, or by combining the velocity transformation with the TV-relation. Related applications can be found in studies \citep{Huang1993, Kumar2022, Song2023, ManzoorHasan2024, Debroeyer2024, Mo2024, Modesti2024}.

These studies highlight an important lesson: existing incompressible wall-modeling techniques can be extended to the simulation of compressible flows by incorporating the compressible laws of the wall. Although algebraic wall models are computationally inexpensive and the control-based wall models are accurate, neither approach has been applied in compressible flows, likely due to the lack of effective compressible transformations and an efficient implementation strategy. From this perspective, a wall model that combines the low cost of algebraic approaches with the accuracy of the control-based models would be highly desireable. Building on recent advances in velocity and temperature transformations, this study aims to extend the control-based approach of \citet{Nicoud2001} to the compressible regime. Specifically, we propose a flux-controlled wall model (FCWM) for near-wall modeling of compressible flows. Compared to the conventional equilibrium wall model, the proposed FCWM achieves higher accuracy in compressible turbulent channel flows without solving the boundary layer equations, thereby reducing computational cost.

The paper is organized as follows. Sec.~\ref{sec:methodology} introduces the methodology that leads to the baseline version of FCWM. Sec.~\ref{sec:improved_wall_model} proposes a near-wall correction to enhance model performance at higher Mach numbers. In Sec.~\ref{sec:application}, the wall model is evaluated in turbulent channel flows across a wide range of Mach and Reynolds numbers. Sec.~\ref{sec:discussion} discusses parameter sensitivity and potential challenges of the model. Finally, conclusion remarks are provided in Sec.~\ref{sec:conclusion}.

\section{\label{sec:methodology}Methodology}
The FCWM consists of three key components: (1) the compressible law of the wall based on velocity and temperature transformations; (2) a feedback flux-control strategy to update the mean wall shear stress and heat flux; and (3) a shifted boundary condition for specifying the local shear stress and heat flux. To demonstrate the idea, we focus on turbulent channel flow. Throughout this study, \( x \), \( y \), and \( z \) denote the streamwise, wall-normal, and spanwise directions, respectively. 
\(\phi\) denotes the filtered quantity. An overline represents Reynolds averaging of \(\phi\) in spatially homogeneous directions and in time, expressed as \( \phi = \bar\phi + \phi^\prime \). A tilde denotes Favre averaging, given by \( \phi = \tilde \phi + \phi^{\prime\prime} \), where \(\tilde\phi = \overline{\rho \phi}/\bar \rho \). The subscript \(w\) denotes wall quantities, and superscript \(+\) indicates normalization by them.

\subsection{\label{sec:control-based_wall_model}Revisiting the control-based wall model by Nicoud et. al}
Different from the conventional ODE-based wall-stress models, \citet{Nicoud2001} proposed to determine the wall shear stress using a control-based approach, which consists of three core steps. First, the plane-averaged differences between the actual and reference velocity profiles in \(u\) and \(w\) at a given \(y-\)plane are defined as:
\begin{subequations}\label{eq:control_wall_model}
  \begin{align}
  \delta_u(y) &= \frac{1}{A} \iint (u - u_{ref}) \, dx \, dz, \label{eq:suboptimal_u} \\
  \delta_w(y) &= \frac{1}{A} \iint (w - w_{ref}) \, dx \, dz, \label{eq:suboptimal_w}
  \end{align}
\end{subequations}
where \(A\) represents the channel area in wall-parallel directions. The reference streamwise velocity \(u_{ref}\) is given by \(u^+_{ref} = \frac{1}{\kappa} \, \log y^+ + C\), and the reference spanwise velocity is \(w_{ref} = 0\) in a fully developed turbulent channel flow. Following Eq.~(\ref{eq:control_wall_model}), a loss function is defined to quantify the mismatch between the computed and reference mean velocity profiles across the domain:
\begin{equation}\label{eq:suboptimal_loss_function}
  J\left(\tau_{w,x}, \tau_{w,z}\right) = \int_{0}^{2h} \left( \delta_u(y)^2 + \delta_w(y)^2 \right) \, dy
  + \frac{\alpha}{A} \iint_{y = 0, 2h} (\tau_{w,x}^2 + \tau_{w,z}^2) \, dx \, dz.
\end{equation}

Here, \(\tau_{w,x}\) and \(\tau_{w,z}\) represent the local shear stresses in the \(x-\) and \(z-\) directions. The second term is introduced to prevent the imposed shear stress from becoming excessively large, thereby avoiding numerical instability \citep{Nicoud2001}. Parameter \(\alpha\) serves to balance the two terms. Finally, the wall shear stresses, \(\tau_{w,x}\) and \(\tau_{w,z}\), can be determined by minimizing \(J\left(\tau_{w,x}, \tau_{w,z}\right)\) and subsequently passed to the outer LES solver as boundary conditions.

Note that Eq.~(\ref{eq:control_wall_model}) requires that the first off-wall cell center is located in the logarithmic layer. In practice, a more realistic reference profile considering the wake region \citep{Smits2024} can also be used. Since the velocity distribution is strongly influenced by the wall shear stress, the first term in Eq.~(\ref{eq:suboptimal_loss_function}) also depends on \(\tau_{w,x}\) and \(\tau_{w,z}\). The primary shortcoming of this approach lies in its computational cost. To obtain the correct \(\tau_{w,x}\) and \(\tau_{w,z}\), gradient-based optimization are performed within each time step of the outer LES, typically requiring approximately 10 iterations \citep{Nicoud2001}. Each iteration involves advancing the state equations and solving the adjoint equations. Consequently, the total computational cost is approximately 20 times larger than that of the algebraic wall model \citep{Nicoud2001}. Although \citet{Templeton2006,Templeton2008} introduced improvements to reduce the cost, the core optimization framework has not been revised. Additionally, this control-based approach was originally designed for incompressible flows. It cannot be directly applied to compressible case.

In the following, we introduce a new flux-control strategy that is based on the recently proposed compressible law of the wall for velocity and temperature distributions. This approach provides the appropriate shear stress and heat flux at the wall, with a computational cost comparable to that of an algebraic wall model.

\subsection{\label{sec:compressible_law_of_the_wall}Compressible law of the wall}
\begin{figure*}[ht]
  \includegraphics[width=0.8\linewidth]{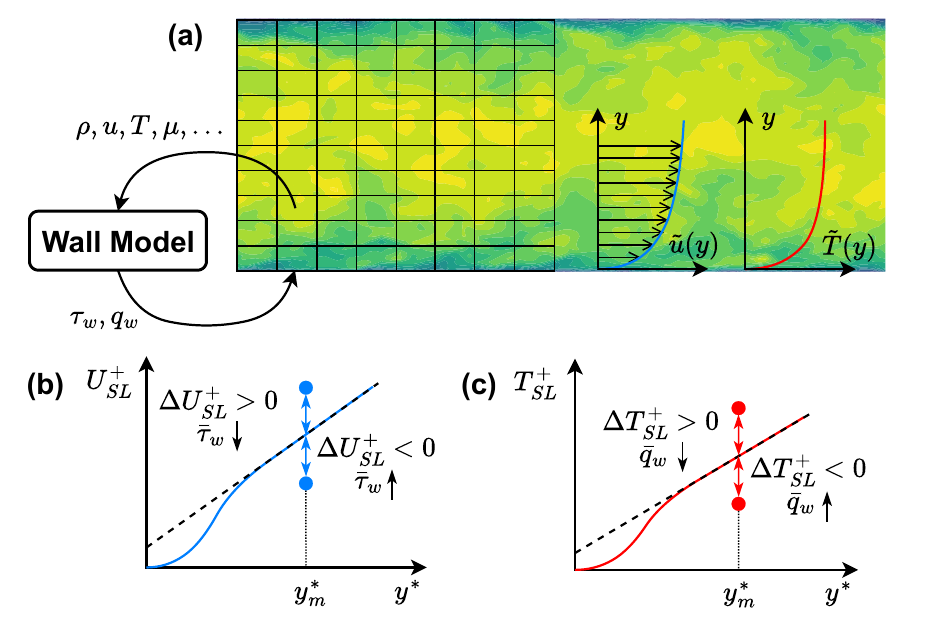}
  \caption{Schematic of the flux-controlled wall model. (a) WMLES setup of compressible wall-bounded turbulent flow. (b) SL-type transformed velocity profile. (c) SL-type transformed temperature profile. When \(\bar \tau_w < \tau_{ref}\), it follows that \(\Delta U^+_{SL} > 0\), and vice versa. Analogously, \(\bar q_w < q_{ref}\) implies \(\Delta T^+_{SL} > 0\), or equivalently \(\bar T(y) > T_{ref}(y)\), and vice versa. Note that the blue and red curves in panels (b, c) represent the transformed velocity and temperature with extended logarithmic profile.}
\label{fig:wall_model_sketch}
\end{figure*}

Fig.~\ref{fig:wall_model_sketch} presents the schematic of FCWM. As illustrated in panel (a), the WMLES employs a uniform coarse grid. An off-wall matching location \(y_m\) is designated for exchanging information between the wall model and the outer LES solver. Flow variables such as \(\rho, u, T\), and \(\mu\) at \(y_m\)---or their profiles below this point---are supplied to the wall model, which in turn provides the wall shear stress (\(\tau_w\)) and heat flux (\(q_w\)) as boundary conditions for the outer LES. Analogous to the incompressible flows, previous studies \citep{Trettel2016, Xu2025b} reveal that the transformed velocity and temperature profiles also present logarithmic behavior in compressible flows, as illustrated in panels (b,c). The reference logarithmic profiles (black dashed lines) are given by:
\begin{subequations}\label{eq:loglaw_Uplus_Tplus}
  \begin{align}
    U^+_{SL} &= \frac{1}{\kappa} \log(y^*) + B, \label{eq:log_law_Uplus}
    \\
    T^+_{SL} &= \frac{Pr_t}{\kappa} \log(y^*) + B_T. \label{eq:log_law_Tplus}
  \end{align}
\end{subequations}

Here, \(U^+_{SL}\) and \(T^+_{SL}\) denote the transformed velocity and temperature. The semi-local wall-normal coordinate is defined as \(y^* = \sqrt{\bar\tau_w \bar\rho} \ y/\bar \mu\), where \(\bar\rho\) and \(\bar\mu\) are the local mean density and dynamic viscosity. \(B\) and \(B_T\) are the intercepts for the velocity and temperature log-laws. \(\kappa\) is the von Kármán constant, and \(Pr_t\) is the turbulent Prandtl number. In this study, \(\kappa = 0.41\) is used. Although recent studies suggest slightly different values \citep{Nagib2008, Lee2015, She2017, Liakopoulos2024}, this choice remains a reasonable estimate within a \(5\%\) error margin \citep{Pope2000}. Analogously, we adopt \(Pr_t = 0.85\), which has been reported to be suitable in the logarithmic region \citep{Bradshaw1995,Lusher2022}.

In the present study, \(U^+_{SL}\) and \(T^+_{SL}\) are computed using the semi-local type (SL-type) velocity and temperature transformation proposed by \citet{Xu2025a, Xu2025b}:
\begin{equation}\label{eq:Uplus_SL}
  U^+_{SL} = \int_0^{u^+} \beta\sqrt{\rho^+} 
  \bigg(1 + \frac{1}{2}\frac{y^+}{\rho^+}\frac{d\rho^+}{dy^+} - \frac{y^+}{\mu^+}\frac{d\mu^+}{dy^+}\bigg) du^+,
\end{equation}
\begin{equation}\label{eq:Tplus_SL}
  T^+_{SL} = \int_0^{\left |\theta^+ \right |}
  \frac{\psi_1}{\left | B_q + \psi_2 {(\gamma - 1)M_\tau^2 u^+} + \psi_3 \right |}
  \sqrt{\rho^+} 
  \bigg(1 + \frac{1}{2}\frac{y^+}{\rho^+}\frac{d\rho^+}{dy^+} - \frac{y^+}{\mu^+}\frac{d\mu^+}{dy^+}\bigg)
  {d \left | \theta^+ \right |},
\end{equation}
with:
\begin{equation}\label{eq:beta_f1f2f3}
  \beta = \frac{l_m}{\kappa y \sqrt{\tau^+_{tot}}},
  \ \ \ 
  \psi_1 = \frac{l_m\sqrt{\tau^+_{tot}}}{\kappa y}, 
  \ \ \ 
  \psi_2 = \tau^+_{tot} + \frac {\tilde u^i_b}{\tilde u} \frac{y}{h},
  \ \ \ 
  \psi_3 = \frac 
  {-\overline{\rho v^{\prime\prime} \frac{1}{2}u^{\prime\prime}_i u^{\prime\prime}_i}}
  {\bar\rho_w u_\tau c_p \tilde T_w}.
\end{equation}

Here, the non-dimensional velocity is defined as \(u^+ = \tilde u/u_\tau\), with the friction velocity given by \(u_\tau = \sqrt{\bar \tau_w/\bar \rho_w}\). The non-dimensional density, viscosity, and temperature difference are defined as \(\rho^+ = \bar\rho/\bar\rho_w\), \(\mu^+ = \bar\mu/\bar\mu_w\), and \(\theta^+ = (\tilde T_w - \tilde T)/{\tilde T_w}\). \(h\) denotes the channel half-height. \(l_m\) is the mixing length. \(\tau^+_{tot} = \tau_{tot}/\bar\tau_w\) represents the normalized total shear stress. In turbulent channel flow, we have \(\tau^+_{tot} = 1 - y/h\). The non-dimensional heat flux is defined as \(B_q = {- \bar q_w}/{(\bar\rho_w c_p u_\tau \tilde T_w)}\) with \(\bar q_w\) denoting heat flux removed from the channel. The friction Mach number is \(M_\tau = {u_\tau}/{\sqrt{\gamma R \tilde T_w}}\), where \(\gamma\) is the ratio of specific heats and \(R\) is the gas constant. \(\tilde u^i_b\) represents the integral bulk velocity, defined as \(\tilde u^i_b = \frac{1}{y}\int_0^y {\tilde u(\eta)}d\eta\).

Note that the velocity transformation in Eq.~(\ref{eq:Uplus_SL}) is a revised form of the transformation originally proposed by \citet{Patel2016} and \citet{Trettel2016}. With the parameters \(\beta\), Eq.~(\ref{eq:Uplus_SL}) yields an extended logarithmic profile in turbulent channel flow compared to the original version \citep{Patel2016, Trettel2016}. For the temperature transformation, the absolute value in Eq.~(\ref{eq:Tplus_SL}) is applied, which is crucial for numerical stability in the FCWM. Regarding \(l_m\), we apply the enhanced mixing length model proposed in our previous study \citep{Xu2025a}, given by:
\begin{subequations}\label{eq:lm}
  \begin{align}
  \frac{l_m}{h} &= 
    \begin{cases}
      \displaystyle \kappa \frac{y}{h}\sqrt{1 - \frac{y}{h}}  & \text{for } y/h \in [0, \eta_{mix}], \\[3ex]
      \displaystyle \frac{K_{mix}(1 - r^{M_{mix}})}{M_{mix}(1+r^2_{core})^{1/4}} \left[1 + \left(\frac{r_{core}}{r}\right)^2 \right]^{1/4}  & \text{for } y/h \in (\eta_{mix}, 1], \\
    \end{cases} \label{eq:enhanced_lm} 
  \\
  \eta_{mix} &= 0.060 + 0.340\exp{(-Re^*_\tau/595)}, \label{eq:eta_mix}
  \\
  K_{mix} &= 0.416 + 0.172 \exp{(-Re^*_\tau/373)}, \label{eq:K_mix}
  \\
  M_{mix} &= 3.104 + 0.871 \exp{(-Re^*_\tau/3144)}, \label{eq:M_mix}
  \end{align}
\end{subequations}
where \(r = 1 - y/h\), \(r_{core} = 0.27\), and \(Re^*_\tau = \sqrt{\bar \tau_w \bar \rho_c} h/\bar \mu_c\) represents the semi-local friction Reynolds number, where the subscript \(c\) denotes quantities at the channel centerline. Our previous studies \citep{Xu2025a, Xu2025b} indicate that applying Eq.~(\ref{eq:lm}) in Eqs.~(\ref{eq:Uplus_SL}) and (\ref{eq:Tplus_SL}) extends the logarithmic behavior in the transformed velocity and temperature, as illustrated in Fig.~\ref{fig:wall_model_sketch} (b) and (c), and further validated by Fig.~\ref{fig:GV2023_SL_Uplus_Tplus}. The extended logarithmic velocity and temperature profiles improve the robustness of the current wall model.

\begin{figure*}[ht]
  \centering
  \begin{tikzpicture}
    \node[anchor=south west, inner sep=0] (image) at (0,0) {\includegraphics[width=\linewidth]{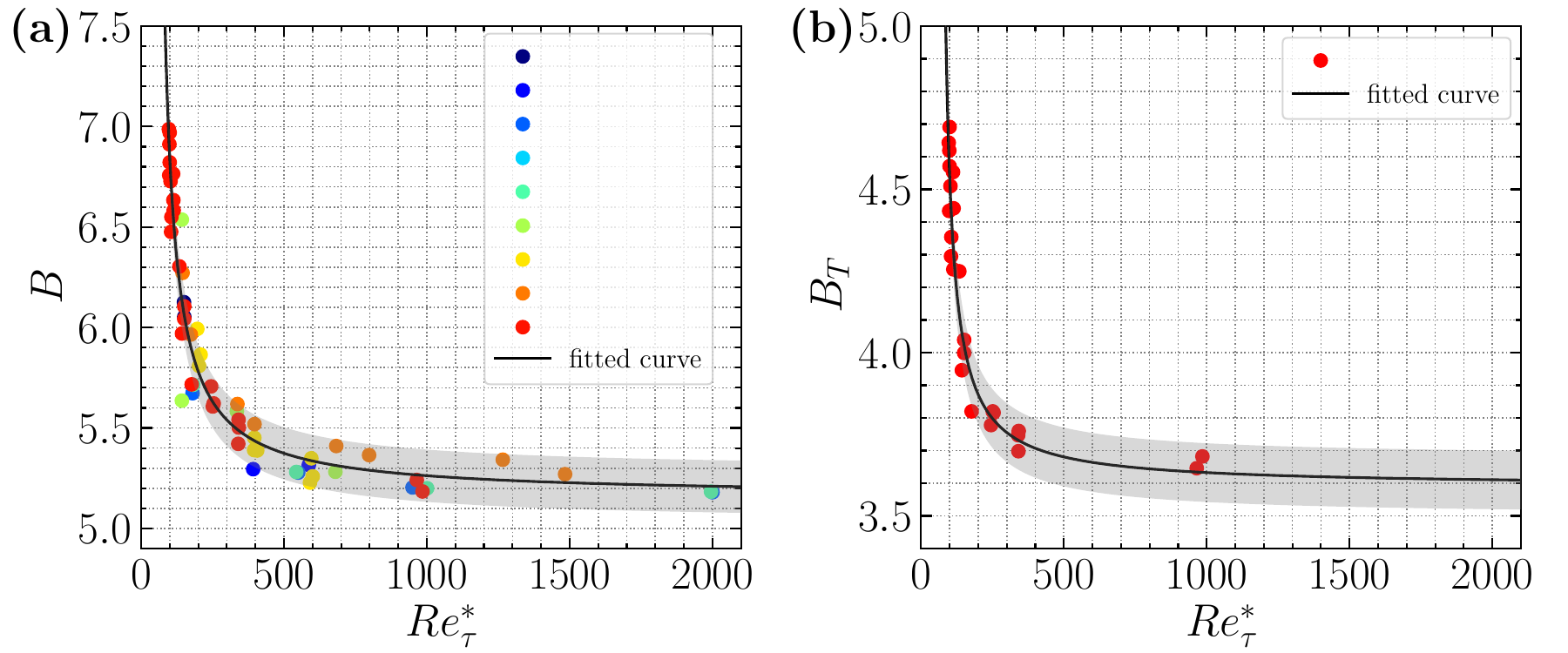}};
    \node[anchor=west, font=\fontsize{6}{11}\selectfont] at (5.8, 6.28) {CKM1995 \citep{Coleman1995}};
    \node[anchor=west, font=\fontsize{6}{11}\selectfont] at (5.8, 5.90) {MKM1999 \citep{Moser1999}};
    \node[anchor=west, font=\fontsize{6}{11}\selectfont] at (5.8, 5.52) {HJ2008 \citep{Hoyas2008}};
    \node[anchor=west, font=\fontsize{6}{11}\selectfont] at (5.8, 5.18) {LJ2014 \citep{Lozano-Duran2014}};
    \node[anchor=west, font=\fontsize{6}{11}\selectfont] at (5.8, 4.80) {LM2015 \citep{Lee2015}};
    \node[anchor=west, font=\fontsize{6}{11}\selectfont] at (5.8, 4.46) {MP2016 \citep{Modesti2016}};
    \node[anchor=west, font=\fontsize{6}{11}\selectfont] at (5.8, 4.10) {TL2016 \citep{Trettel2016}};
    \node[anchor=west, font=\fontsize{6}{11}\selectfont] at (5.8, 3.75) {YH2020 \citep{Yao2020}};
    \node[anchor=west, font=\fontsize{6}{11}\selectfont] at (5.8, 3.40) {GV2023 \citep{Gerolymos2023, Gerolymos2024a, Gerolymos2024b}};
    \node[anchor=west, font=\fontsize{6}{11}\selectfont] at (14.0, 6.20) {GV2023 \citep{Gerolymos2023, Gerolymos2024a, Gerolymos2024b}};
  \end{tikzpicture}
  \caption{Dependence of log-law intercepts on \(Re^*_\tau\) for (a) transformed velocity and (b) transformed temperature. The shaded areas represent an error margin of \(\pm \, rms\) for \(B\) and \(B_T\). Note that \(Re^*_\tau = Re_\tau\) for incompressible flows.}
\label{fig:loglaw_B_BT}
\end{figure*}
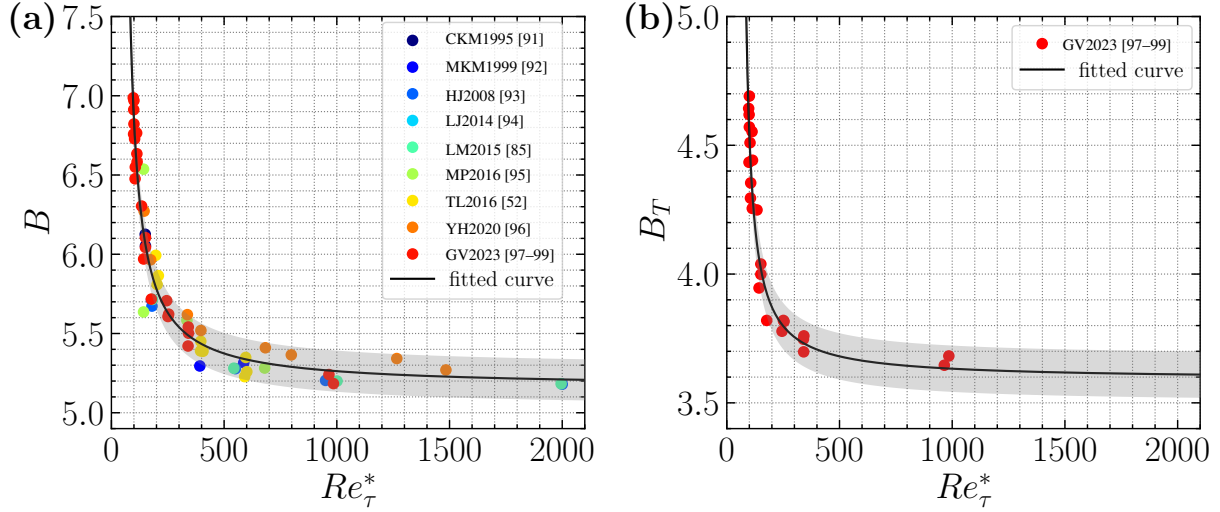

WMLES is typically applied at high Reynolds numbers, where the log-law intercepts in Eq.(\ref{eq:loglaw_Uplus_Tplus}) can be treated as constants. However, at lower Reynolds numbers, variations in \(B\) and \(B_T\) are often observed in both incompressible \citep{Nagib2008} and compressible flows \citep{Brun2008, Trettel2016, Griffin2021, Chen2022a, Cheng2024a}. \citet{Hasan2023} showed that incorporating intrinsic compressibility into the transformation by \citet{Trettel2016} reduces this shift. Nevertheless, the shift is not completely eliminated in the classical isothermal wall configuration of compressible turbulent channel flow. For practical applicability, we fit \(B\) and \(B_T\) to Reynolds number using available DNS data \citep{Coleman1995, Moser1999, Hoyas2008, Lozano-Duran2014, Lee2015, Modesti2016, Trettel2016, Yao2020, Gerolymos2023, Gerolymos2024a, Gerolymos2024b}, as shown in Fig.~\ref{fig:loglaw_B_BT}. For \(B\), both incompressible and compressible DNS data are used. In the case of \(B_T\), however, most of the publicly available DNS data do not contain the high-order statistic required in Eq.~(\ref{eq:Tplus_SL}). Therefore, only the DNS data from \citet{Gerolymos2023, Gerolymos2024a, Gerolymos2024b} are applied. The fitted results are:
\begin{subequations}\label{eq:loglaw_B_BT}
  \begin{align}
  B &= \frac{98}{Re^*_\tau - 42} + 5.16\,  \label{eq:loglaw_B},
  \\
  B_T &= \frac{40} {Re^*_\tau - 58} + 3.59.  \label{eq:loglaw_BT}
  \end{align}
\end{subequations}

Based on Eq.~(\ref{eq:loglaw_B_BT}), the log-law intercepts asymptotically approach \(B = 5.16\) and \(B_T = 3.59\) at sufficiently high Reynolds numbers, assuming a von Kármán constant \(\kappa = 0.41\) and turbulent Prandtl number \(Pr_t = 0.85\). In Fig.~\ref{fig:loglaw_B_BT}, error margin of \(\pm \, rms\) for \(B\) and \(B_T\) are also shown, which cover most of the datasets for \(Re^*_\tau > 200\). As most DNS data for compressible turbulent channel flows are obtained at relatively low Reynolds numbers, and compressibility effects further reduce the effective Reynolds number \(Re^*_\tau\) \citep{Trettel2016}, the corrections in Eq.~(\ref{eq:loglaw_B_BT}) are necessary and will be applied throughout this study.

\subsection{\label{sec:feedback_flux_control_strategy}Feedback flux-control strategy}
Since the log-law is inherently a statistical feature in wall-bounded turbulence, \(U^+_{SL}\) and \(T^+_{SL}\) usually do not align exactly with the reference logarithmic profiles at each individual time step. Instead, slight deviations are often observed. This is more evident in WMLES where coarse grid is applied. Thus, optimizing Eq.~(\ref{eq:suboptimal_loss_function}) at every time step, as done by \citet{Nicoud2001}, imposes overly strict requirements. This is especially true for compressible flows, where three variables, \(\tau_{w,x}(x,z), \tau_{w,z}(x,z)\), and \(q_w(x,z)\), must be optimized simultaneously. Furthermore, the velocity and temperature transformations introduce additional complexity. Applying the same approaches as \citet{Nicoud2001} and \citet{Templeton2006, Templeton2008} would significantly increase computational cost. Therefore, two key modifications are introduced:
\begin{enumerate}
  \renewcommand{\labelenumi}{(\theenumi)}
  \item Instead of matching the entire transformed velocity and temperature profiles to the reference log-law, only data at the matching location \(y_m\) are considered.
  \item Perfect matching of the log-law at the matching location within each time step is not required. Instead, a feedback control is employed to statistically guide the transformed velocity and temperature towards the reference log-law.
\end{enumerate}

To this end, at each time step \(n\), the differences between the computed and reference values of the transformed velocity and temperature at \(y^*_m\) are computed as follows:
\begin{subequations}\label{eq:delta_Uplus_Tplus}
  \begin{align}
    \left. \Delta U^+_{SL} \right|_{y^*_m}^{n} = \left. U_{SL}^{+} \right|_{y^*_m}^{n} - \left. U_{SL}^{+,loglaw} \right|_{y^*_m}^{n},  \label{eq:delta_Uplus}
  \\
  \left. \Delta T^+_{SL} \right|_{y^*_m}^{n} = \left. T_{SL}^{+} \right|_{y^*_m}^{n} - \left. T_{SL}^{+,loglaw} \right|_{y^*_m}^{n}.  \label{eq:delta_Tplus}
  \end{align}
\end{subequations}

Note that the matching location \(y_m\) does not necessarily coincide with a cell center. If it does not, linear interpolation is applied in Eq.~(\ref{eq:delta_Uplus_Tplus}) to evaluate \(\left. U_{SL}^{+} \right|_{y^*_m}^{n}\) and \(\left. T_{SL}^{+} \right|_{y^*_m}^{n}\). 

For a fully developed turbulent channel flow, when \(\bar \tau_w < \tau_{ref}\), we usually have \(\Delta U^+_{SL} > 0\), indicating that the shear stress should be increased, and vice versa. Similarly, when \(\bar q_w < q_{ref}\), we usually find \(\Delta T^+_{SL} > 0\), or equivalently \(\bar T(y) > T_{ref}(y)\), suggesting that the heat flux should be increased, and vice versa. Here, \(\tau_{ref}\) and \(q_{ref}\) represent the true shear stress and heat flux, respectively. In this study, we propose to leverage these observations in a transient, inverse manner to adjust the shear stress and heat flux: 
\begin{enumerate}
  \renewcommand{\labelenumi}{(\theenumi)}
  \item \( ( \left. \Delta U_{SL}^{+} \right|_{y^*_m}^{n} > 0, \left. \Delta T^+_{SL} \right|_{y^*_m}^{n} > 0 ) \): increase both \(\bar\tau_w\) and \(\bar q_w\).
  \item \( ( \left. \Delta U_{SL}^{+} \right|_{y^*_m}^{n} < 0, \left. \Delta T^+_{SL} \right|_{y^*_m}^{n} < 0 ) \): decrease both \(\bar\tau_w\) and \(\bar q_w\).
  \item \( ( \left. \Delta U_{SL}^{+} \right|_{y^*_m}^{n} > 0, \left. \Delta T^+_{SL} \right|_{y^*_m}^{n} < 0 ) \): increase \(\bar\tau_w\) and decrease \(\bar q_w\).
  \item \( ( \left. \Delta U_{SL}^{+} \right|_{y^*_m}^{n} < 0, \left. \Delta T^+_{SL} \right|_{y^*_m}^{n} > 0 ) \): decrease \(\bar\tau_w\) and increase \(\bar q_w\).
\end{enumerate}

Here, \(\left. \Delta U_{SL}^{+} \right|_{y^*_m}\) and \(\left. \Delta T^+_{SL} \right|_{y^*_m}\) serve as the "loss function", analogous to \(J\) in Eq.~(\ref{eq:suboptimal_loss_function}). The control objective is to determine the correct shear stress and heat flux such that \(\left. \Delta U_{SL}^{+} \right|_{y^*_m} \approx 0\) and \(\left. \Delta T^+_{SL} \right|_{y^*_m} \approx 0\) in a statistical sense. Inspired by the methodology of \citet{Bae2022}, we propose the following flux-control strategy:
\begin{subequations}\label{eq:update_tau_q}
  \begin{align}
    \bar\tau_w^{n} &= a_{\tau}^n \, \bar\tau_w^{n-1} \ \text{with} \ a_{\tau}^n = 1 + \lambda_\tau\tanh\,\big(\left.{\Delta U_{SL}^{+}}\right|_{y^*_m}^n \big),  \label{eq:update_tau}
  \\
  \bar q_w^{n} &= a_q^n \, \bar q_w^{n-1}  \ \text{with} \ a_q^n = 1 + \lambda_q\tanh\,\big(\left.{\Delta T_{SL}^{+}} \right|_{y^*_m}^n \big).  \label{eq:update_q}
  \end{align}
\end{subequations}

Here, \(\lambda_\tau\) and \(\lambda_q\) serve as wall-flux relaxation coefficients that regulate the temporal evolution of the shear stress and heat flux, respectively. Our experiences show that \(\lambda_\tau = \lambda_q = 0.01 \sim 0.08\) is suitable to guarantee stability and accuracy. In contrast, values of \(\lambda_\tau\) and \( \lambda_q\) greater than 0.1 tend to produce noticeable discrepancies in the temperature distribution. 

\subsection{\label{sec:Shifted_boundary_condition}Shifted boundary condition}
In practical simulation, the local wall shear stress and heat flux must be specified at each wall-adjacent cell face. Several methods have been proposed to impose the local boundary conditions \citep{Schumann1975, Groetzbach1987,Piomelli1989}. In this study, we adopt the shifted boundary condition proposed by \citet{Piomelli1989}, which correlates the instantaneous wall shear stress to the velocity at a downstream location in an off-wall plane. Using this approach, the local shear stress and heat flux are determined by:
\begin{subequations}\label{eq:shifted_boundary_condition}
\begin{align}
  \tau_{w, x}^n(x, z) &= \frac{u(x + \Delta_s, y_1, z)}{\bar u(y_1)} \ \bar\tau_w^n, \label{eq:tau_x}
  \\
  \tau_{w, z}^n(x, z) &= \frac{w(x + \Delta_s, y_1, z)}{\bar u(y_1)} \ \bar\tau_w^n, \label{eq:tau_w}
  \\
  q_w^n(x, z) &= \frac{T(x + \Delta_s, y_1, z) - \bar T_w}{\bar T(y_1) - \bar T_w}\ \bar q_w^n.  \label{eq:q_w}
\end{align}
\end{subequations}

Here, \(y_1\) denotes the first off-wall cell center. $\Delta_s$ is a streamwise displacement approximately given by \( \Delta_s = y_1 \ \text{cot}(8^\circ) \) for \( 30 < y^+_1 < 50 \), and \( \Delta_s = y_1 \ \text{cot}(18^\circ) \) for larger \(y^+_1\) \citep{Piomelli-Balaras}. In this formulation, the local shear stress and heat flux are assumed to be proportional to the corresponding velocity components and temperature difference at a downstream location in the \(y_1-\)plane. For the wall-normal velocity, no-penetration condition is imposed, i.e.,  \(v(x, z) = 0\).

In summary, the FCWM integrates the compressible law of the wall, a feedback flux-control strategy, and a shifted boundary condition. Instead of seeking the optimal values of \(\bar\tau_w\) and \(\bar q_w\) at every time step, the model incrementally adjusts the shear stress and heat flux to statistically converge to correct values over time. As a result, the transformed velocity and temperature at the matching location \(y^*_m\) align with the reference logarithmic profiles. The wall model introduced above is referred to as the baseline flux-controlled wall model, denoted as "FCWM-base".

\subsection{\label{sec:preliminary_evaluation}Preliminary evaluation of FCWM-base}
In this subsection, we evaluate the performance of the proposed baseline wall model. All results in this study are obtained using the computational fluid dynamics (CFD) solver JAX-Fluids \citep{Bezgin2023, Bezgin2025a}, which has been validated in previous studies \citep{Bezgin2023, Bezgin2025a, Bezgin2025b, Feng2024, Xu2025b}. We perform wall-modeled implicit LES using the Adaptive Local Deconvolution Method (ALDM), as introduced in \citet{Adams2004, Hickel2006, Hickel2007}, and \citet{Hickel2014}. Unlike explicit LES, implicit LES filters the flow variables through finite-volume discretization, with the SGS effects incorporated via the numerical scheme and flux function, without introducing additional non-linear terms. Previous studies have also validated the application of ALDM for WRLES \citep{Doehring2018, Doehring2021} and WMLES \citep{Chen2010, Chen2011, Chen2014}. The governing equations are given by:
\begin{equation}\label{eq:mass_conservation}
  \frac{\partial \rho}{\partial t} + \frac{\partial {\rho u_j}}{\partial x_j} = 0,
\end{equation}
\begin{equation}\label{eq:momentum_conservation}
  \frac{\partial \rho u_i}{\partial t} 
  + \frac{\partial {\rho u_j u_i}}{\partial x_j} 
  = -\frac{\partial p}{\partial x_i} + \frac{\partial \tau_{ij}}{\partial x_j}
  + f_1 \delta_{i1},
\end{equation}
\begin{equation}\label{eq:energy_conservation}
  \frac{\partial }{\partial t} \bigg [\rho \bigg(c_v T + \frac{u_i u_i}{2}\bigg) \bigg]
  + \frac{\partial }{\partial x_j} \bigg [\bigg(\rho c_v T + \frac{\rho u_i u_i}{2} + p \bigg) u_j \bigg]
  = \frac{\partial \tau_{ij} u_i}{\partial x_j}
  - \frac{\partial q_j}{\partial x_j}
  + f_1 u_1,
\end{equation}
with the viscous stress \(\tau_{ij}\) and heat flux vector \(q_j\) given by:
\begin{equation}\label{eq:tau_ij}
\tau_{ij} = \mu \bigg(\frac{\partial u_i}{\partial x_j} 
+ \frac{\partial u_j}{\partial x_i} 
- \frac{2}{3}\frac{\partial u_k}{\partial x_k}\delta_{ij}\bigg), \ \ q_j = - k \frac{\partial T}{\partial x_j}.
\end{equation}

Here, \( x \), \( y \), and \( z \) denote the streamwise, wall-normal, and spanwise directions, respectively. The velocity components in these directions are represented by \( u_i(i=1,2,3) \). \(t\) is the time, \(\rho\) is the fluids density, \(p\) is the pressure, \(c_v\) is the specific heat capacity at constant volume. \(\delta_{ij}\) is the Kronecker delta. \(\mu\) and \(k\) represent the molecular dynamic viscosity and thermal conductivity. To close the governing equations, the state equation \( p = \rho R T \) is invoked, where \(R\) is the idea gas constant. The specific heat capacity at constant volume is given by \(c_v = 1/(\gamma - 1) R\), with a constant ratio of specific heats \( \gamma = 1.4 \). The flow is driven by a uniform body force \( f_1 \) in streamwise direction to maintain a constant mass flow rate.

The wall model is employed to provide the appropriate wall shear stress and heat flux at the wall. To account for the near-wall behavior of the SGS stress, the following damping function is proposed for ALDM \citep{Hickel2007}:
\begin{equation}\label{eq:ALDM_inner_damping}
  f_{VD}^{inner} = \left[1 - \exp\left(-\left(\frac{l^+_w}{A^+}\right)^d\right) \right]^s, \text{where} \,\, A^+ = 50, d = 3, s = 1/3
\end{equation}

Here, \(l^+_w\) denotes the inner-scaled wall-normal distance. This damping function is successfully implemented in the WRLES of turbulent channel flow by \citet{Hickel2007}. However, when applied in WMLES, two shortcomings are observed. First, Eq.~(\ref{eq:ALDM_inner_damping}) does not capture the correct asymptotical behavior in the near-wall region. A more appropriate choice is \(ds = 3\) \citep{Hickel2007, Piomelli1993}, which is consistent with the damping function form introduced in \citet{Balaras1996}. Second, the damping function is no longer physically meaningful as the size of LES grid is very large in terms of the viscous length scale \citep{Chen2014}, i.e., \(y^+_1\) is comparable or larger than \(A^+\). In light of these limitations, and based on our experiences, the following damping function is applied in the present study:
\begin{equation}\label{eq:ALDM_outer_damping}
  f_{VD}^{outer} = 1 - \exp\left[-\left(\frac{l_w}{A}\right)^3\right], \text{where}\, A = \max\left(0.08, \frac{5}{3} M_\tau - 0.02\right).
\end{equation}

Here the outer-scaled wall-normal distance \(l_w = y/h\) is applied in place of \(l^+_w\). The dependence of \(A\) on \(M_\tau\) reflects the compressibility effects \citep{Hasan2023}. A detailed discussion of this dependence is beyond the scope of the present study.

Two representative flow conditions for compressible turbulent channel flows are considered: \(M_b = 0.74, Re_b = 21{,}092\) and \(M_b = 1.57, Re_b = 25{,}216\), where the bulk Mach number is defined as \(M_b = U_b/\sqrt{\gamma R \tilde T_w}\), and the bulk Reynolds number as \(Re_b = \rho_b U_b h/\bar \mu_w\). Here, the bulk density and velocity are computed as \(\rho_b = \frac{1}{2h}\int_0^{2h} \bar\rho dy\) and \(U_b = \int_0^{2h} \overline{\rho u} dy/(2\rho_b h)\), respectively. These flow conditions match the DNS of \citet{Gerolymos2023, Gerolymos2024a, Gerolymos2024b}. The computational domain is set to \(L_x \times L_y \times L_z = 2\pi h \times 2h \times \pi h\), which is demonstrated by \citet{Lozano-Duran2014} to be sufficient to produce correct one-point statistics.

Unlike WRLES and DNS, the grid resolution in WMLES is typically measured using outer scaling, such as the boundary layer thickness or the channel half-height. \citet{Larsson2016} recommend to use grid spacing of \(\Delta x/h \approx 0.08\), \(\Delta y/h \approx 0.02 \sim 0.05\), and \(\Delta z/h \approx 0.05\), respectively. Unless otherwise stated, a uniform grid of \(N_x \times N_y \times N_z = 104 \times 40 \times 64\) is applied throughout this study, corresponding to \(\Delta x/h \approx 0.06\), \(\Delta y/h \approx 0.05\), and \(\Delta z/h \approx 0.05\), respectively. Given the results in \citet{Kawai2012} and analogous to \citet{Griffin2023}, \(y_m = 0.3 h\) is employed. During the simulation, a fourth order central finite-difference scheme is used for spatial discretization of dissipative fluxes, while a third-order Runge-Kutta (RK3) method is employed for time integration with a Courant-Friedrichs-Lewy (CFL) number
of 0.8. Throughout this study, the Prandtl number is set to \(Pr = 0.7\), and the dynamic viscosity follows a power law relationship, given by \(\mu/\mu_w = (T/T_w)^{0.7}\). Periodic boundary conditions are applied in the streamwise and spanwise directions, while the bottom and top walls are maintained at fixed temperature of \(T_w = 1.0\). The proposed wall model provides shear stress and heat flux imposed on the walls. In addition, the trapezoid method is applied to numerically integrate Eqs.~(\ref{eq:Uplus_SL}) and (\ref{eq:Tplus_SL}). The simulation is initialized from the converged simulation on a \(32\times32\times32\) grid without employing a wall model.

\begin{figure*}[ht]
  \includegraphics[width=0.8\linewidth]{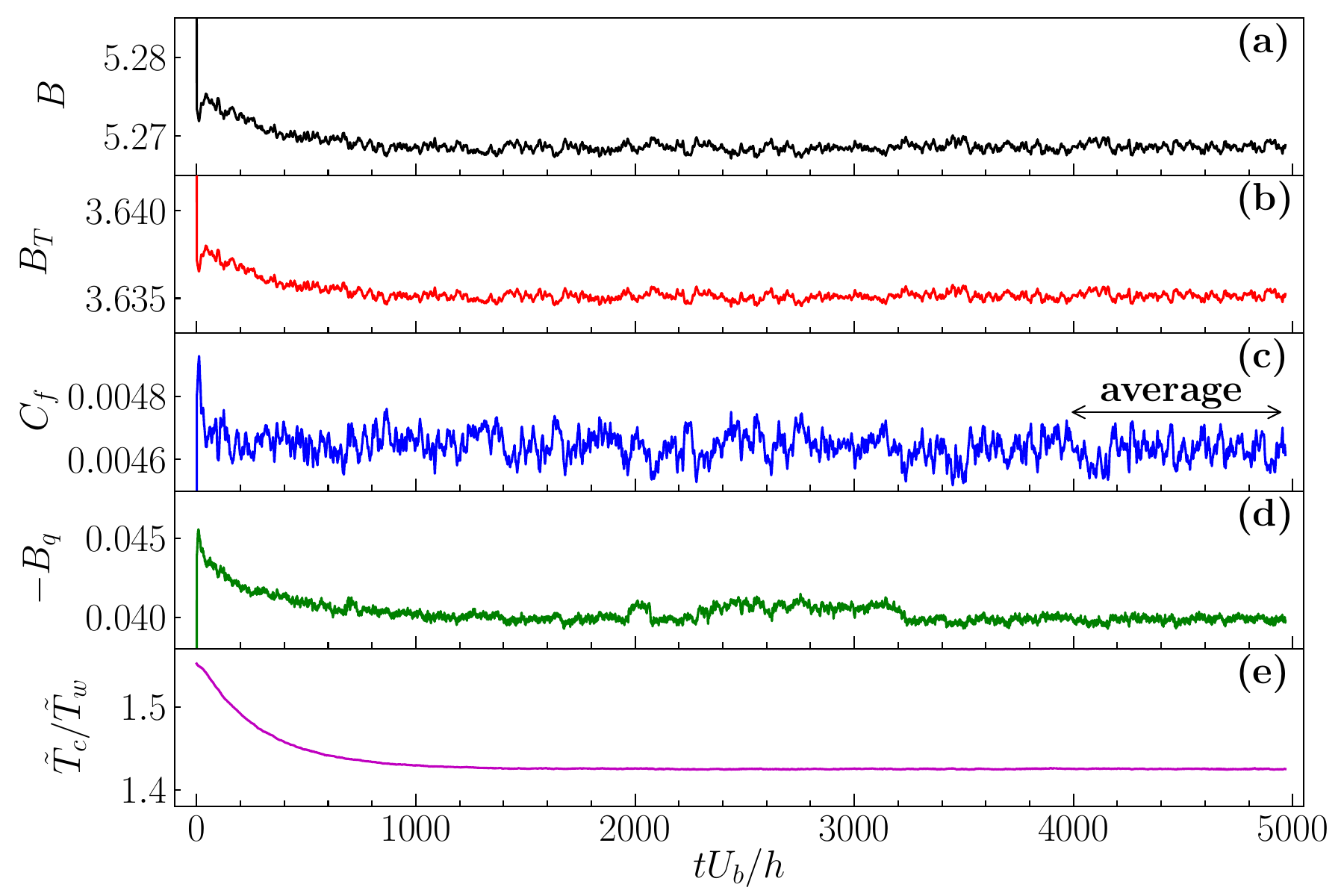}
  \caption{Evolution of (a) log-law intercept \(B\) for transformed velocity, (b) log-law intercept \(B_T\) for transformed temperature, (c) friction coefficient \(C_f\), (d) non-dimensional heat flux \(B_q\), and (e) mean centerline temperature \(\tilde T_c/\tilde T_w\) for WMLES of compressible turbulent channel flow at \(M_b = 1.57\) and \(Re_b = 25{,}216\), using a uniform grid of \(104 \times 40 \times 64\).}
\label{fig:time_series}
\end{figure*}

Fig.~\ref{fig:time_series} presents the variation of \(B\), \(B_T\), \(C_f\), \(-B_q\), and \(\tilde T_c/\tilde T_w\) during the simulation of the case at \(M_b = 1.57, Re_b = 25{,}216\) using \(f^{outer}_{VD}\). Here, \(C_f\) denotes the friction coefficient, defined as \(C_f = \bar \tau_w/(\frac{1}{2}\rho_b U_b^2)\). It is observed that both \(B\) and \(B_T\) exhibit fluctuations but statistically converge to the values prescribed by Eq.~(\ref{eq:loglaw_B_BT}). A similar behavior is observed for \(C_f\), \(-B_q\), and \(\tilde T_c/\tilde T_w\). Finally, the flow statistics are computed via temporal and wall-parallel averaging over 500 snapshots, covering approximately 40 turnover times (\(\Delta t \approx 40 h/u_\tau\)), or equivalently about \(1{,}000 h/U_b\), as shown in panel (c). 
\begin{figure*}
  \includegraphics[width=1\linewidth]{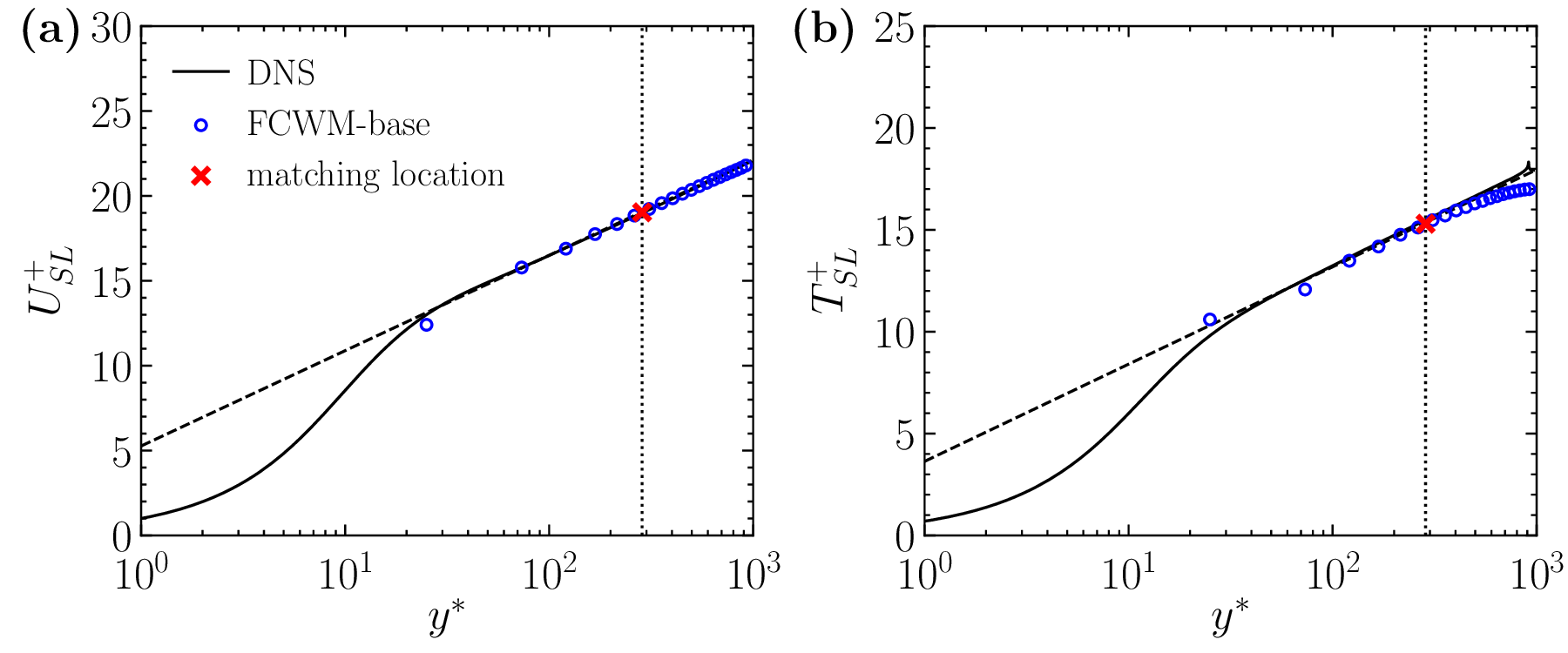}
  \caption{Converged profiles of (a) \(U^+_{SL}\) and (b) \(T^+_{SL}\) from WMLES of compressible turbulent channel flow at \(M_b = 1.57\) and \(Re_b = 25{,}216\), using FCWM-base and \(f^{outer}_{VD}\). The black dashed lines: \(U^+_{SL} = \frac{1}{\kappa} \log (y^*) + 5.27\) and \(T^+_{SL} = \frac{Pr_t}{\kappa} \log (y^*) + 3.64\). The red crosses {$\boldsymbol{\times}$} mark the matching location (\(y_m = 0.3h\)). DNS data from \citet{Gerolymos2023, Gerolymos2024a, Gerolymos2024b} are included for comparison.}
\label{fig:GV2023_SL_Uplus_Tplus}
\end{figure*}

The transformed velocity and temperature profiles for the case at \(M_b = 1.57\) and \(Re_b = 25{,}216\) are presented in Fig.~\ref{fig:GV2023_SL_Uplus_Tplus}. Note that the vertical dotted lines mark the matching location \(y^*_m\), with the two red crosses indicating \((y^*_m, \left. U_{SL}^{+} \right|_{y^*_m})\) and \((y^*_m, \left. T_{SL}^{+} \right|_{y^*_m})\). As shown, both points lie on the reference logarithmic profiles (black dashed lines), indicating \(\left. \Delta U_{SL}^{+} \right|_{y^*_m} = 0\) and \(\left. \Delta T_{SL}^{+} \right|_{y^*_m} = 0\). A similar state is realized for the case at \(M_b = 0.74\) and \(Re_b = 21{,}094\) (not shown here).

Although only \(U_{SL}^{+}\) and \(T_{SL}^{+}\) at \(y^*_m\) are directly controlled, the overall profiles largely adhere to the log-law in the entire outer layer. However, it should be noted that these alignments only reflect convergence of the control strategies defined in Eqs.~(\ref{eq:delta_Uplus_Tplus}) and (\ref{eq:update_tau_q}), and do not imply the overall accuracy of the WMLES, which is different from the incompressible case \citep{Nicoud2001}.

\begin{figure*}[ht]
  \includegraphics[width=1\linewidth]{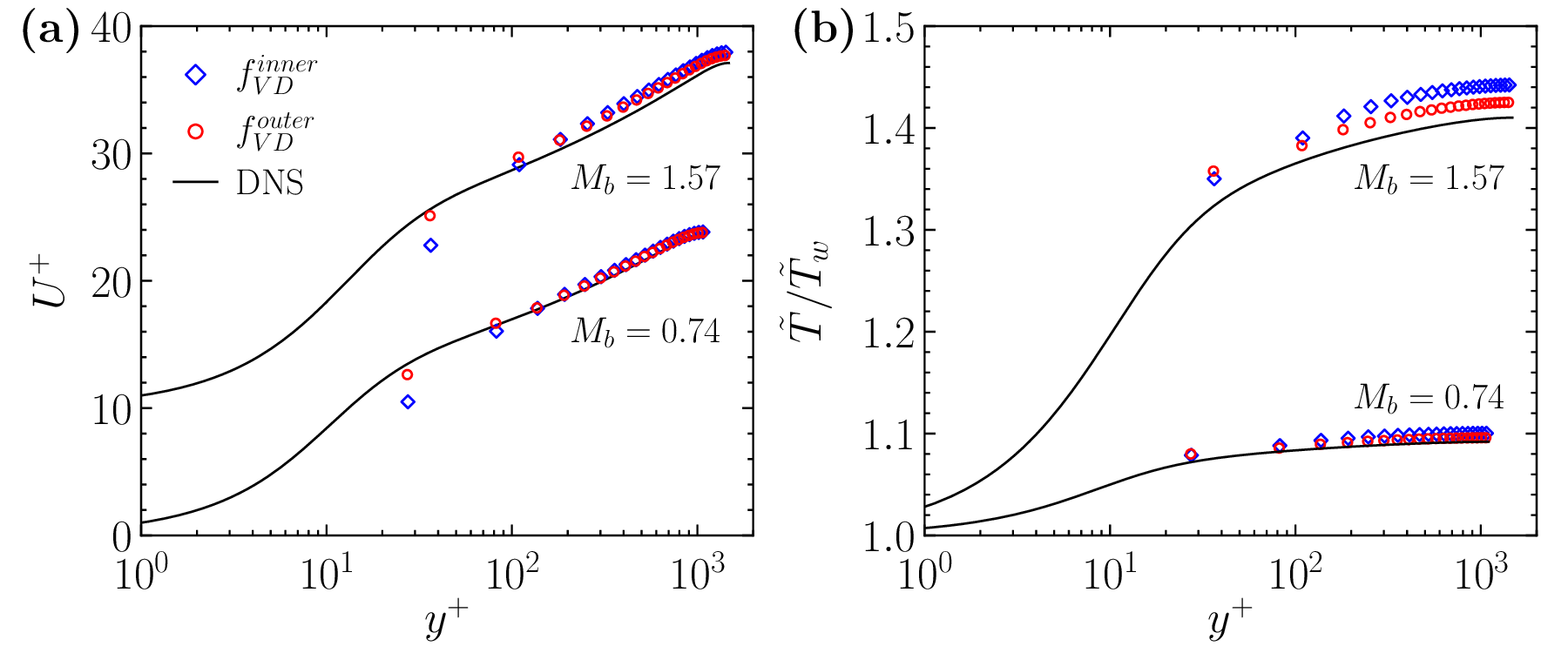}
  \caption{Profiles of (a) velocity and (b) temperature in WMLES of compressible turbulent channel flow at \(M_b = 0.74, Re_b = 21{,}092\) and \(M_b = 1.57, Re_b = 25{,}216\), using \(f^{inner}_{VD}\) and \(f^{outer}_{VD}\) defined in Eqs.~(\ref{eq:ALDM_inner_damping}) and (\ref{eq:ALDM_outer_damping}). DNS data from \citet{Gerolymos2023, Gerolymos2024a, Gerolymos2024b} are included for comparison.}
\label{fig:GV2023_Uplus_T_Tw_fVD}
\end{figure*}

To assess the accuracy, the untransformed velocity and temperature profiles using both \(f^{inner}_{VD}\) and \(f^{outer}_{VD}\) are presented in Fig.~\ref{fig:GV2023_Uplus_T_Tw_fVD}. For the case at \(M_b = 0.74\) and \(Re_b = 21{,}092\), both \(U^+\) and \(\tilde T/\tilde T_w\) show good agreement with DNS results. However, slight discrepancies are observed for the case at \(M_b = 1.57\) and \(Re_b = 25{,}216\). Compared to \(f^{inner}_{VD}\), the \(f^{outer}_{VD}\) yields closer agreements with DNS data. Particularly, it reduces the discrepancy in velocity at the first off-wall cell center, as shown in panel (a). This is primarily due to the reduced equivalent SGS stress in the near-wall region.

Apart from the velocity and temperature profiles, WMLES should also provide reasonable predictions of \(C_f\), \(B_q\), and \(\tilde T_c\), which are important for engineering applications. The relative error of these quantities is computed as follows:
\begin{equation}\label{eq:error_Cf}
  \epsilon_{C_f} = \frac{C_f^{wm} - C_f^{DNS}}{C_f^{DNS}} \times 100\%,
\end{equation}
\begin{equation}\label{eq:error_Bq}
  \epsilon_{B_q} = \frac{-B_q^{wm} + B_q^{DNS}}{-B_q^{DNS}} \times 100\%,
\end{equation}
\begin{equation}\label{eq:error_Tc_Tw}
  \epsilon_{T_c} = \frac{(\tilde T_c/\tilde T_w)^{wm} - (\tilde T_c/\tilde T_w)^{DNS}}{(\tilde T_c/\tilde T_w)^{DNS}} \times 100\%.
\end{equation}

Using \(f^{outer}_{VD}\), the relative errors of the two cases are:
\begin{itemize}
  \item \(M_b = 0.74, Re_b = 21{,}092\): \(\epsilon_{C_f} = -2.12\%, \epsilon_{B_q} = -1.00\%, \text{and} \, \epsilon_{T_c} = -0.37\%\).
  \item \(M_b = 1.57, Re_b = 25{,}216\): \(\epsilon_{C_f} = -5.52\%, \epsilon_{B_q} = -2.72\%, \text{and} \, \epsilon_{T_c} = 1.06\%\).
\end{itemize}

Using \(f^{inner}_{VD}\) yields better prediction of \(C_f\), and worse prediction of \(B_q\) and \(T_c\). Additional details are provided in Table~\ref{table:cases_of_WMLES}.

Based on the above results, two conclusions can be drawn: (1) The revised damping function \(f^{outer}_{VD}\) outperforms the original \(f^{inner}_{VD}\). (2) FCWM-base performs well at relatively low Mach numbers, but its accuracy reduces as the Mach number increases. For the case at \(M_b = 1.57, Re_b = 25{,}216\), the relative errors in \(C_f\), \(B_q\), and \(\tilde T_c/\tilde T_w\) on a representative grid resolution remain within acceptable limits. However, at higher Mach numbers, its reliability cannot be guaranteed. Although increasing wall-normal grid resolution improves the accuracy, it conflicts with the primary objective of reducing computational cost by using a coarse mesh in WMLES. In light of these limitations, further improvements are introduced in the following section.

\section{\label{sec:improved_wall_model} Improved wall model}
To improve the accuracy at higher Mach numbers, we propose a near-wall correction to the proposed wall model, yielding significant improvements in the computed velocity and temperature profiles, as well as in \(\epsilon_{C_f}\), \(\epsilon_{B_q}\), and \(\epsilon_{T_c}\).

\subsection{\label{sec:near_wall_correction} Near-wall correction}
In typical WMLES, the near-wall region is under-resolved due to the application of coarse grid. This compromises the accuracy of FCWM for compressible flow simulations, as the density and viscosity profiles, along with their gradients, cannot be accurately captured in this region. As a result, integration errors become inevitable in the velocity and temperature transformations given in Eqs.~(\ref{eq:Uplus_SL}) and (\ref{eq:Tplus_SL}). To illustrate this, we isolate and rewrite the terms related to density and viscosity as follows:
\begin{equation}\label{eq:g_rhomu}
  G^{\rho\mu} = 
  \sqrt{\rho^+} 
  \left(1 + \frac{1}{2}\frac{y^+}{\rho^+}\frac{d\rho^+}{dy^+} 
  - \frac{y^+}{\mu^+}\frac{d\mu^+}{dy^+}\right).
\end{equation}

\begin{figure*}
  \includegraphics[width=1.0\linewidth]{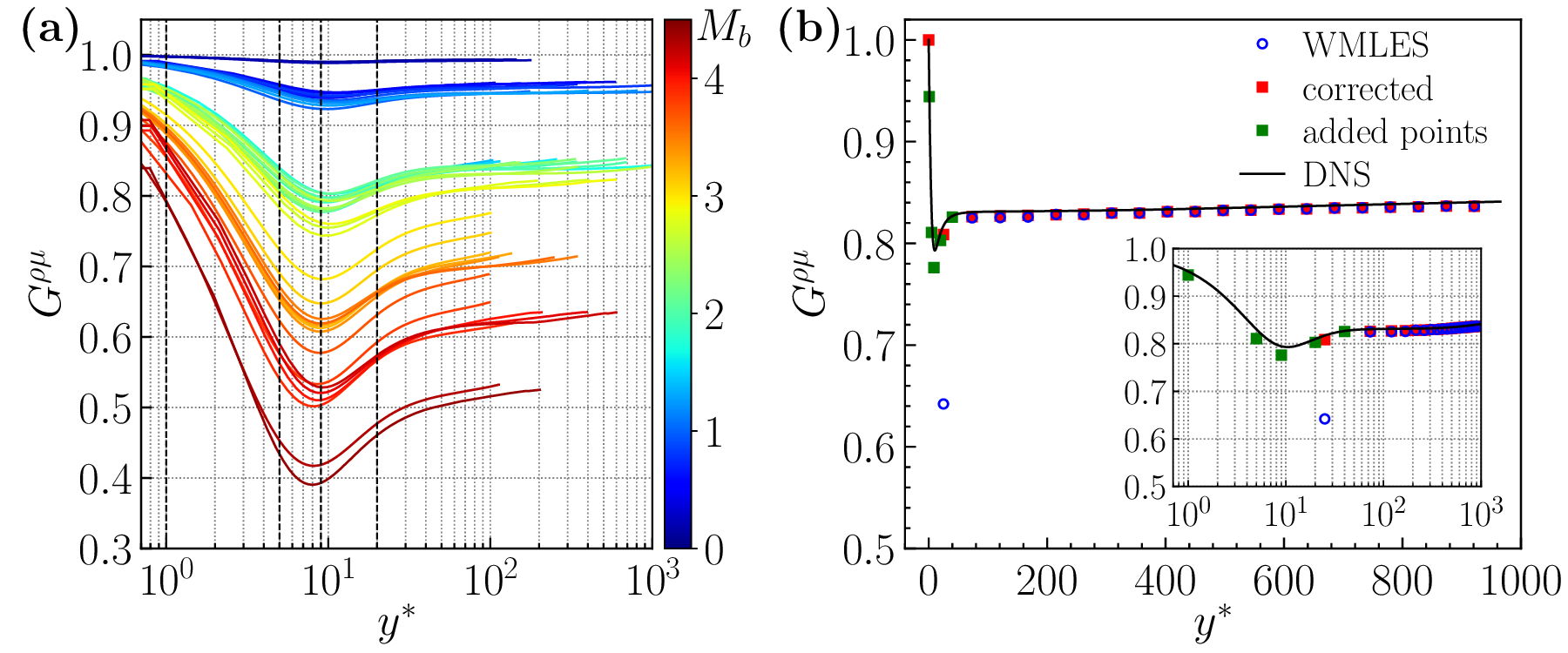}
  \caption{(a) \(G^{\rho\mu}\) profiles at various Mach and Reynolds numbers, computed from DNS data \citep{Modesti2016,Trettel2016,Yao2020,Gerolymos2023, Gerolymos2024a, Gerolymos2024b}. (b) \(G^{\rho\mu}\) profile from WMLES at \(M_b = 1.57\) and \(Re_b = 25{,}216\). The DNS profile for this case is computed from data of \citet{Gerolymos2023, Gerolymos2024a, Gerolymos2024b}.}
\label{fig:g1g2_profile}
\end{figure*}

For incompressible flows, the fluid properties are constant, yielding \(G^{\rho\mu} = 1.0\), and the transformations reduce to their incompressible forms. As Mach number increases, this no longer holds, as shown in Fig.~\ref{fig:g1g2_profile}. In panel (a), the \(G^{\rho\mu}\) distributions at various Mach and Reynolds numbers computed from DNS data \citep{Coleman1995,Modesti2016,Trettel2016,Yao2020, Gerolymos2023, Zhu2025b} are presented with respect to \(y^*\). Panel (b) presents the corresponding profile from a WMLES at \(M_b = 1.57\) and \(Re_b = 25{,}216\) using FCWM-base. Two observations can be made:
\begin{enumerate}
  \renewcommand{\labelenumi}{(\theenumi)}
  \item \(G^{\rho\mu}\) presents Mach-number-dependent behavior, and shows negligible sensitivity to Reynolds number. At very low Mach numbers, the \(G^{\rho\mu}\) profile approaches the incompressible limit. As the Mach number increases, it systematically deviates from 1.0---dropping significantly in the near-wall region, with a minimum at \(y^* \approx 9\). Beyond \(y^* \approx 40\), it approaches an approximate plateau. 
  \item In WMLES, the computed \(G^{\rho\mu}\) generally agrees with DNS in the main flow. However, the accuracy is reduced near the wall, particularly at the first off-wall cell center. This underprediction of \(G^{\rho\mu}\) in this region is directly connected to the overprediction of temperature profile in Fig.~\ref{fig:GV2023_Uplus_T_Tw_fVD} (b).
\end{enumerate}

\begin{figure*}
  \includegraphics[width=1.0\linewidth]{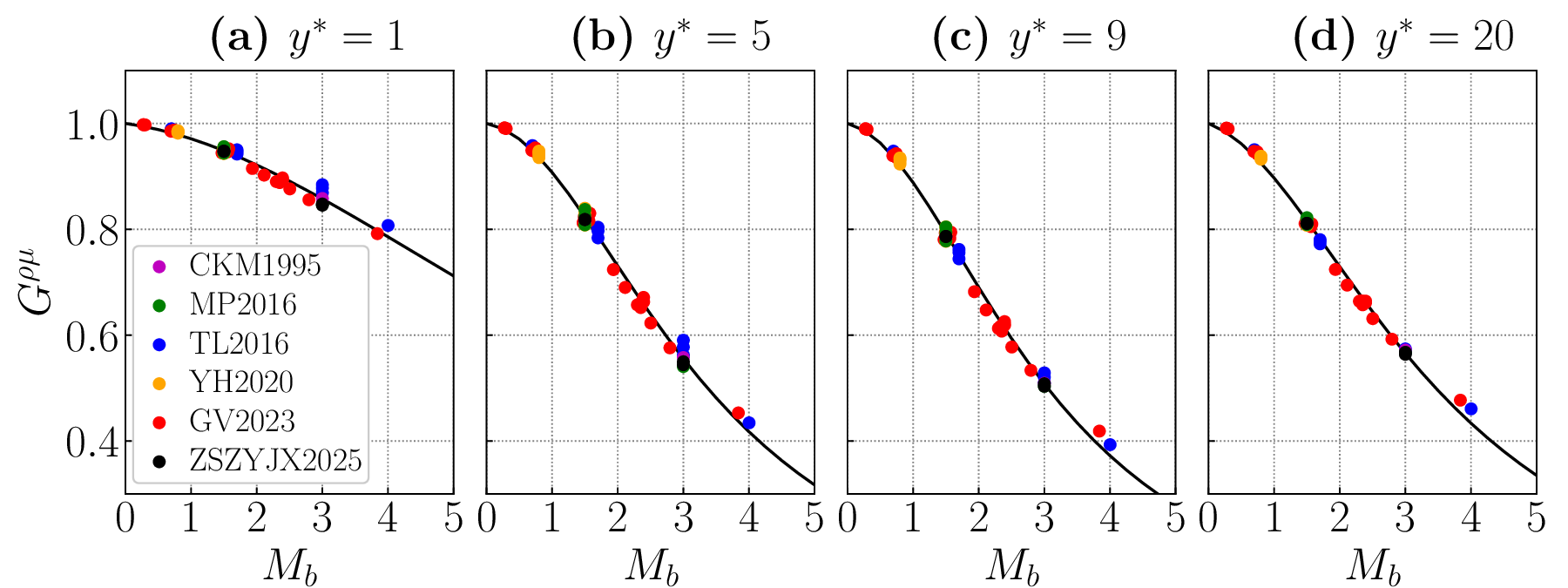}
  \caption{Dependence of \(G^{\rho\mu}\) on Mach numbers at (a) \(y^* = 1\), (b) \(y^* = 5\), (c) \(y^* = 9\), and (d) \(y^* = 20\). Curves represent the fitted correlations computed from DNS datasets \citep{Coleman1995,Modesti2016,Trettel2016,Yao2020,Zhu2025b,Gerolymos2023}.}
\label{fig:g1g2_1_5_9_20}
\end{figure*}

Based on these observations, we propose manually supplementing the missing \(G^{\rho\mu}\) in the near-wall region and applying corrections to \(G^{\rho\mu}\) below the matching location. This can be achieved through the following steps:

\textbf{Step 1}: The values of \(G^{\rho\mu}\) at \(y^* =\) 1, 5, 9, and 20 are supplemented, as indicated by the black dashed lines in Fig.~\ref{fig:g1g2_profile}(a). \(G^{\rho\mu}\) at these points strongly depend on the Mach number, as shown in Fig.~\ref{fig:g1g2_1_5_9_20}. By fitting DNS data from open literatures \citep{Coleman1995,Modesti2016, Trettel2016, Yao2020, Gerolymos2023, Zhu2025b}, the following correlations are obtained:
\begin{subequations}\label{eq:fitted_g1g2}
  \begin{align}
    \left. G^{\rho\mu} \right|_{y^*_m = 1} &= 1/(1 + 0.017 M_b + 0.013 M_b^2),  \label{eq:g1g2_1}
    \\
    \left. G^{\rho\mu} \right|_{y^*_m = 5} &= 1/(1 + 0.019 M_b + 0.082 M_b^2),  \label{eq:g1g2_5}
    \\
    \left. G^{\rho\mu} \right|_{y^*_m = 9} &= 1/(1 + 0.027 M_b + 0.099 M_b^2),  \label{eq:g1g2_9}
    \\
    \left. G^{\rho\mu} \right|_{y^*_m = 20} &= 1/(1 + 0.044 M_b + 0.071 M_b^2).  \label{eq:g1g2_20}
  \end{align}
\end{subequations}

\textbf{Step 2}: The value of \(G^{\rho\mu}\) at \(y^* = 40\) is estimated from results in the WMLES main flow using linear interpolation between reference points at \(y/h = 0.5\) and \(y/h = 0.8\).

\textbf{Step 3}: The \(G^{\rho\mu}\) profile on the LES grid for \(0 \leq y^* \leq y^*_m\) is corrected by linear interpolation using values at \(y^* = 1, 5, 9, 20, 40\), and \(y/h = 0.5\).

Following the above steps, the corrected \(G^{\rho\mu}\) profile for WMLES of the case at \(M_b = 1.57\) and \(Re_b = 25{,}216\) is illustrated in Fig.~\ref{fig:g1g2_profile}(b). The filled red squares represent the corrected values at the LES cell centers, while the filled green squares indicate the supplemented values at \(y^* = 1, 5, 9, 20, 40\). The corrected \(G^{\rho\mu}\) shows significantly improved agreements with DNS results. For consistency, each term in Eqs.~(\ref{eq:Uplus_SL}) and (\ref{eq:Tplus_SL}) should also include values at \(y^* = 1, 5, 9, 20, 40\), which are obtained by linear interpolation in the present study.

\begin{figure*}
  \includegraphics[width=0.75\linewidth]{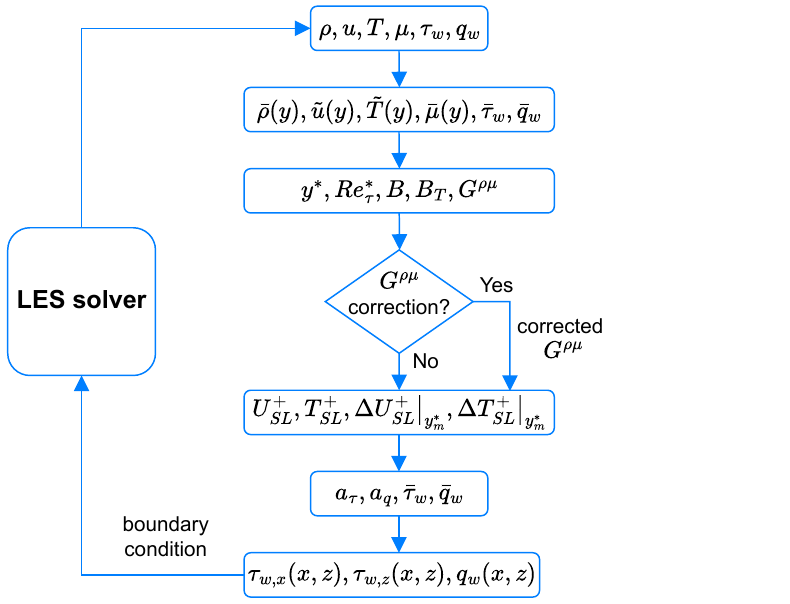}
  \caption{Schematic of the flux-controlled wall model.}
\label{fig:wall_model_schematic}
\end{figure*}

Combining the baseline wall model introduced in Sec.~\ref{sec:methodology} with the corrections presented above, the complete schematic of the wall model is illustrated in Fig.~\ref{fig:wall_model_schematic}, with implementation details provided in Algorithm \ref{alg:wall_model_implementation}. Two versions of the flux-controlled wall model are obtained:
\begin{enumerate}
  \renewcommand{\labelenumi}{(\theenumi)}
  \item FCWM-base, without the \(G^{\rho\mu}\) correction.
  \item FCWM-G, with the \(G^{\rho\mu}\) correction.
\end{enumerate}

Note that when doing the numerical integration, the value of \(G^{\rho\mu}\) between \(y^* = 1\) and \(y^* = 5\) is observed to vary linearly with respect to \(\log y^*\), as indicated in Fig.~\ref{fig:g1g2_profile} (a).

\begin{algorithm}[ht]
  \caption{FCWM implementation from time step \(n\) to \(n + 1\)}
  \label{alg:wall_model_implementation}

  \DontPrintSemicolon
  \small

  Start from end of time step \(n\) \;
  Obtain 3-D flow field: \(\rho, u, v, w, T, \mu\), and 2-D wall flux: \(\tau_w, q_w\)\;
  Average in wall-parallel direction to obtain \(\bar\rho(y), \tilde u(y), \tilde T(y), \bar\tau_w, \bar q_w, \bar\rho_w\)\;  
  Compute \(y^*, Re^*_\tau\), and obtain \(B, B_T\) using Eq.~(\ref{eq:loglaw_B_BT})\;  
  Compute \(l_m, \beta, \psi_1, \psi_2, \psi_3\) using Eqs.~(\ref{eq:beta_f1f2f3}) and (\ref{eq:lm})\;  
  Compute \(G^{\rho\mu}\) profile using Eq.~(\ref{eq:g_rhomu})\;  

  \If{\({G^{\rho\mu}}\)\textnormal{-correction is applied}}{
    Compute the supplementary values at \(y^* = 1, 5, 9, 20, 40\):\;
    for \(G^{\rho\mu}\), follow steps 1--3 in Sec.~\ref{sec:near_wall_correction}\;
    for \(\bar\rho(y), \tilde u(y), \tilde T(y), l_m, \beta, \psi_1, \psi_2, \psi_3\), use linear interpolation\;
  }

  Compute \(U^+_{SL}\) and \(T^+_{SL}\) profiles using Eqs.~(\ref{eq:Uplus_SL}) and (\ref{eq:Tplus_SL})\;  
  Evaluate computed \(U^+_{SL}\) and \(T^+_{SL}\) at the matching location \(y^*_m\) by linear interpolation\;
  Evaluate reference \(U^+_{SL}, T^+_{SL}\) at matching location \(y^*_m\) using Eq.~(\ref{eq:loglaw_Uplus_Tplus})\;
  Evaluate \(\Delta U^+_{SL}\) and \(\Delta T^+_{SL}\) at matching location \(y^*_m\) using Eq.~(\ref{eq:delta_Uplus_Tplus})\;  
  Compute new mean shear stress and heat flux: \(\bar \tau_w, \bar q_w\) using Eq.~(\ref{eq:update_tau_q})\;  
  Compute local wall shear stress and heat flux \(\tau_w (x,z), q_w (x,z)\) using Eq.~(\ref{eq:shifted_boundary_condition})\;  
  Supply \(\tau_w (x,z)\) and \(q_w (x,z)\) as boundary conditions to the outer LES solver\;
  Proceed to time step \(n + 1\)\;
\end{algorithm}

\subsection{\label{sec:performance_comparison} Performance comparison}
\begin{figure*}
  \includegraphics[width=1.0\linewidth]{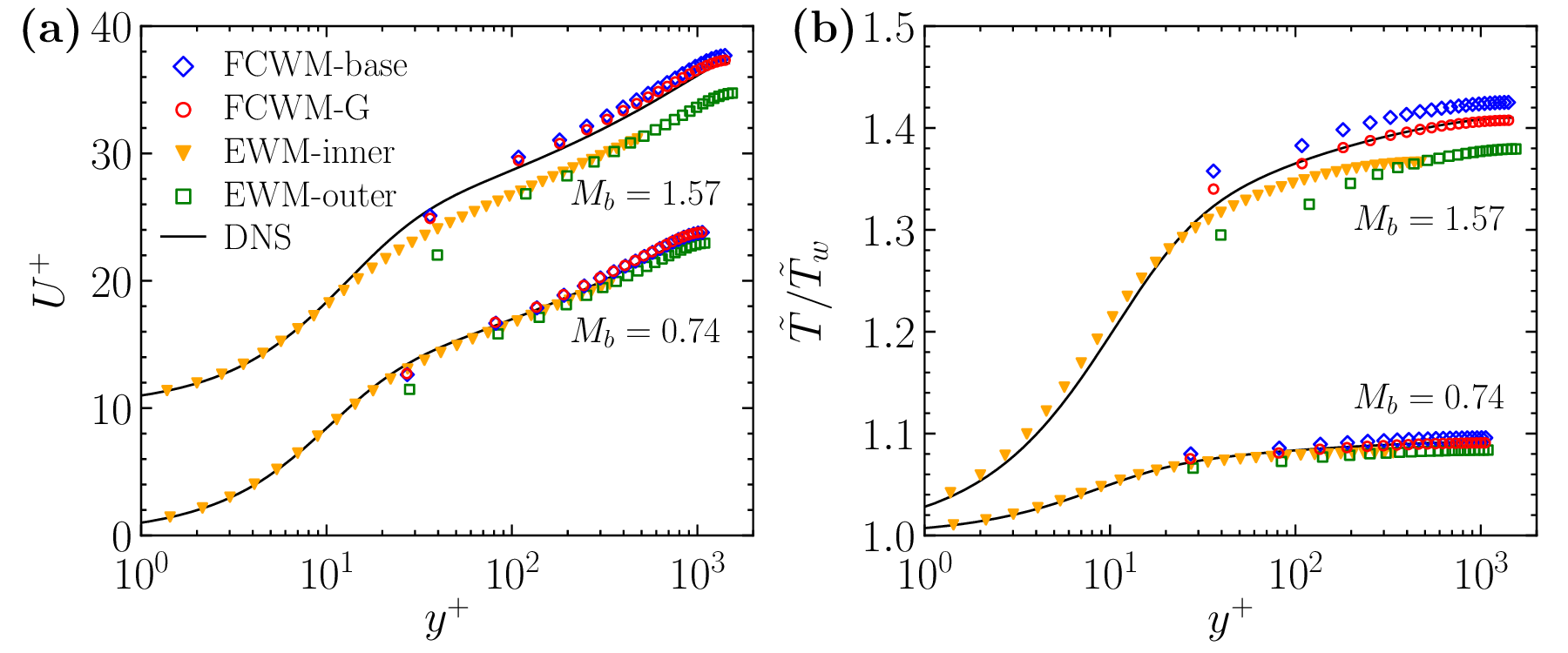}
  \caption{Velocity (a) and temperature (b) profiles in compressible turbulent channel flow at \(M_b = 0.74, Re_b = 21{,}092\) and \(M_b = 1.57, Re_b = 25{,}216\). The \(U^+\) profile for the second case is shifted upward by 10 units. DNS data from \citet{Gerolymos2023, Gerolymos2024a, Gerolymos2024b} are included for comparison.}
\label{fig:GV2023_Uplus_T_Tw}
\end{figure*}

To evaluate the performance of above correction, we perform WMLES under the same conditions as in Sec.~\ref{sec:preliminary_evaluation}, using FCWM-G and \(f^{outer}_{VD}\). The computed velocity and temperature profiles are presented in Fig.~\ref{fig:GV2023_Uplus_T_Tw}. For comparison, results obtained from the classical EWM are also included. Details about this wall model and its validation are provided in the Appendix.

As shown, for the case at \(M_b = 0.74, Re_b = 21{,}092\), all the three wall models provide results in agreement with the DNS data. The classical EMW results in slightly lower \(U^+\) and \(\tilde T/\tilde T\) in the outer solution. For the case at \(M_b = 1.57, Re_b = 25{,}216\), the \(G^{\rho\mu}\) correction significantly improves the accuracy of both \(U^+\) and \(\tilde T/\tilde T_w\) profiles. On the contrary, the EMW results present evident discrepancies in the entire outer layer. Given the nearly constant wall-normal pressure and the ideal gas assumption, a well-predicted temperature profile also implies a reliable density distribution, which is therefore not shown here. The prediction errors of \(\epsilon_{C_f}\), \(\epsilon_{B_q}\) and \(\epsilon_{T_c}\) by FCWM-G are as follows:
\begin{itemize}
  \item \(M_b = 0.74, Re_b = 21{,}092: \epsilon_{C_f} = -2.95\%, \epsilon_{B_q} = -1.28\%, \text{and} \, \epsilon_{T_c} = -0.09\%\).
  \item \(M_b = 1.57, Re_b = 25{,}216: \epsilon_{C_f} = -4.09\%, \epsilon_{B_q} = -1.34\%, \text{and} \, \epsilon_{T_c} = -0.14\%\).
\end{itemize}

Compared to FCWM-base, FCWM-G yields better prediction for all three quantities at \(M_b = 1.57, Re_b = 25{,}216\). Regarding the case at \(M_b = 0.74, Re_b = 21{,}092\), slightly larger discrepancies in \(\epsilon_{C_f}\) and \(\epsilon_{B_q}\) are observed. As will be seen in Sec.~\ref{sec:application}, this is case-dependent, and does not diminish the overall improvements of FCWM-G across a broader ranges of flow conditions. The relative errors of the EWM are considerably larger (not shown here). Detailed comparisons are presented in table~\ref{table:cases_of_WMLES} and Fig.~\ref{fig:error_Cf_Bq_TcTw}.

\section{\label{sec:application}Application}
This section evaluates the performance of the proposed wall model across a broader range of Mach and Reynolds numbers. Two types of flow conditions are considered: quasi-incompressible and compressible turbulent channel flows. For the quasi-incompressible case, we assess the computed mean velocity profile, Reynolds shear stress and friction coefficient \(C_f\), showing that the proposed wall model appropriately reduces to the incompressible limit as the Mach number decreases. For the compressible flows, we focus on velocity and temperature profiles, friction coefficient \(C_f\), non-dimensional heat flux \(B_q\), temperature ratio \(\tilde T_c/\tilde T_w\), and turbulent statistics, thereby directly demonstrating the effectiveness of the proposed wall model under compressible conditions.

\subsection{\label{sec:incompressible_flow}Quasi-incompressible turbulent channel flow}
For quasi-incompressible turbulent channel flow, we examine cases with \(M_b =0.1\) and friction Reynolds number ranging from \(Re_\tau = 180 \sim 10{,}000\), where \( Re_\tau = \sqrt{\bar \tau_w \bar \rho_w} h/\bar \mu_w \). The considered \(Re_\tau\) values are consistent with those used in the DNS studies \citep{Lozano-Duran2014, Lee2015, Yamamoto2018,Oberlack2021}. 

The computational domain for all cases is set to \(2\pi h \times 2h \times \pi h\), and a uniform grid of \(80 \times 30 \times 50\) is used. The matching location is \(y_m = 0.3h\). Since the temperature variation is negligible at such low Mach number, the heat transfer is neglected by applying a source term in the energy equation to maintain a constant temperature matching the wall. This yields nearly uniform density and viscosity fields. Consequently, only the baseline FCWM-base is used: the shear stress is prescribed using Eq.~(\ref{eq:Uplus_SL}), while the heat flux is computed using Fourier's law without applying Eq.~(\ref{eq:Tplus_SL}). This simplification does not compromise the accuracy under quasi-incompressible conditions. The damping function \(f^{outer}_{VD}\) is applied, and other computational settings follow those described in Sec.~\ref{sec:preliminary_evaluation}. Results of the WMLES are provided in Table~\ref{table:quasi_incompressible_WMLES}. Note that only \(M_b\) and \(Re_b\) are prescribed as the input parameters. Quantities such as \(Re_\tau\) and \(C_f\) are simulation results and reflect the accuracy of the simulation.
\begin{table}[ht]
  \caption{\label{table:quasi_incompressible_WMLES} WMLES results for quasi-incompressible turbulent channel flow. The computational domain for all cases is set to \(2\pi h \times 2h \times \pi h\), with a uniform grid of \(80 \times 30 \times 50\). Each case is labeled by its bulk Mach number and friction Reynolds number. For example, "M0.1Retau1000" denotes \(M_b = 0.1\) and \(Re_\tau = 1{,}000\). The two values in each parenthesis correspond to results from WMLES and DNS, with DNS results shown in bold.}.
  \begin{ruledtabular}
  \footnotesize
  \begin{tabular}{l c c c c c c}
    Case & DNS reference & \(M_b\) & \(Re_b\) & \( Re_\tau \) & \( C_f(\times 10^{-3}) \) \\[3pt]
    \hline
    M0.1Retau180   & \citet{Lee2015}          & 0.1 & 2{,}857   & (181, \textbf{182})     & (8.016, \textbf{8.123})  \\
    M0.1Retau550   & \citet{Lee2015}          & 0.1 & 10{,}000  & (543, \textbf{543})     & (5.899, \textbf{5.908})  \\
    M0.1Retau1000  & \citet{Lee2015}          & 0.1 & 20{,}000  & (998, \textbf{1001})    & (4.982, \textbf{5.005})  \\
    M0.1Retau2000  & \citet{Lee2015}          & 0.1 & 43{,}650  & (1998, \textbf{1995})   & (4.190, \textbf{4.210})  \\
    M0.1Retau4200  & \citet{Lozano-Duran2014} & 0.1 & 98{,}304  & (4169, \textbf{4179})   & (3.597, \textbf{3.614})  \\
    M0.1Retau5200  & \citet{Lee2015}          & 0.1 & 125{,}000 & (5193, \textbf{5186})   & (3.452, \textbf{3.442})  \\
    M0.1Retau8000  & \citet{Yamamoto2018}     & 0.1 & 200{,}400 & (7980, \textbf{8016})   & (3.172, \textbf{3.200})  \\
    M0.1Retau10000 & \citet{Oberlack2021}     & 0.1 & 257{,}143 & (10017, \textbf{10049}) & (3.035, \textbf{3.050})  \\    
  \end{tabular}
  \end{ruledtabular}
\end{table}

\begin{figure*}
  \includegraphics[width=\linewidth]{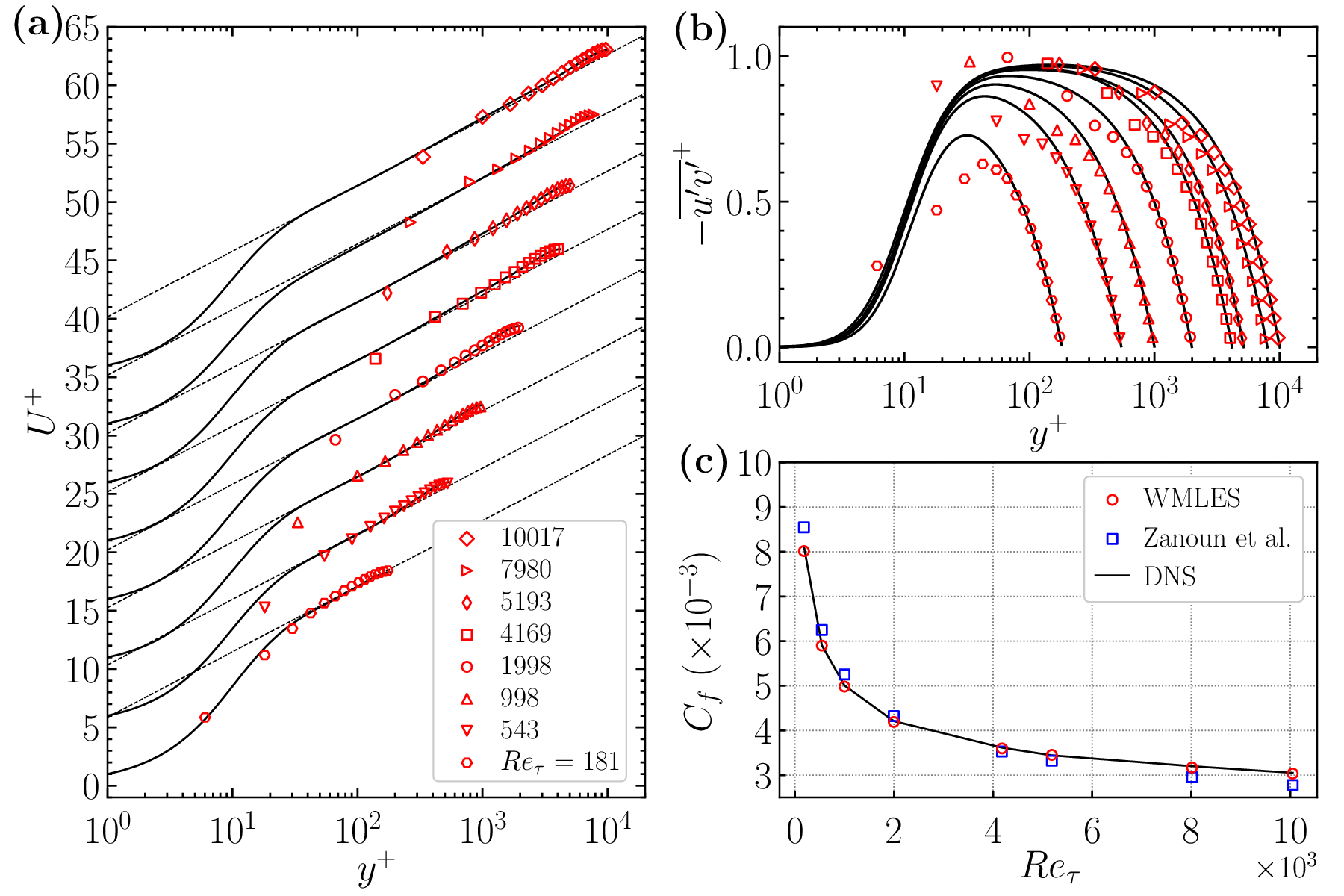}
  \caption{Results of quasi-incompressible turbulent channel flows simulations using the FCWM-base: (a) velocity, (b) Reynolds shear stress, and (c) friction coefficients. The legend in (b) is the same as (a). Solid lines represent DNS data cited in table~\ref{table:quasi_incompressible_WMLES}.}
\label{fig:incomp_uplus_yplus_180_10k}
\end{figure*}

Fig.~\ref{fig:incomp_uplus_yplus_180_10k} presents the computed velocity, Reynolds shear stress, and relative errors. For comparison, the friction coefficients computed from DNS and the empirical relation by \citet{Zanoun2009} are also presented in panel (c), where
\begin{equation}\label{eq:Cf_Zanoun}
  C_f = 0.0743 Re_m^{-0.25} \ \ \text{with} \ Re_m = \frac{\rho_b U_b 2h}{\mu}
\end{equation}

It is evident that the computed velocity profiles show good agreement with DNS results across the broad range of Reynolds numbers. Notably, the LLM commonly observed in many previous WMLES studies \citep{Kawai2012, Yang2017, Maejima2024} is absent for the current wall model. For Reynolds shear stress, the wall model yields satisfactory predictions in the main flow, though noticeable discrepancies appear in the first few off-wall cells. The friction coefficient also align with both DNS data and the empirical correlation. The maximum \(\epsilon_{C_f}\) remains below \(1\%\), except for the case at \(M_b = 0.1\) and \(Re_\tau = 180\), which is expected due to low-Reynolds-number effects \citep{Modesti2016, Griffin2023}. These results confirm that the proposed wall model performs well for quasi-incompressible turbulent channel flows.

\subsection{\label{sec:compressible_flow}Compressible turbulent channel flow}
In addition to the cases at \(M_b = 0.74, Re_b = 21{,}092\) and \(M_b = 1.57, Re_b = 25{,}216\) presented in Sec.~\ref{sec:preliminary_evaluation} and \ref{sec:performance_comparison}, a broader range of Mach and Reynolds numbers is considered for compressible flows, spanning from \(M_b = 0.8\) to \(4.0\) and \(Re_b = 7{,}667\) to \(34{,}000\). These flow conditions cover subsonic to supersonic regimes and moderate to relatively high Reynolds numbers, and are consistent with previous DNS studies \citep{Trettel2016,Modesti2016, Yao2020, Gerolymos2023}. The damping function \(f^{outer}_{VD}\) is applied, and other computational settings follow those described in Sec.~\ref{sec:preliminary_evaluation}. In addition to FCWM-base and FCWM-G, results using EWM are also included in this section for comparison. Additional details of the WMLES are listed in Table~\ref{table:cases_of_WMLES}. Note that, for present compressible turbulent channel flows, only \(M_b\) and \(Re_b\) are prescribed as the input parameters. Quantities such as \(C_f, B_q\), and \(\tilde T_c/\tilde T_w\) are simulation results and reflect the accuracy of the simulation.

\begin{table}[ht]
  \caption{\label{table:cases_of_WMLES} WMLES of compressible turbulent channel flows. The computational domain for all cases is set to \(2\pi h \times 2h \times \pi h\) with a uniform grid of \(104 \times 40 \times 64\). Each case is labeled by its bulk Mach and Reynolds numbers. For example, "M0.74Re21092" denotes \(M_b = 0.74\) and \(Re_b = 21{,}092\). The values in each parenthesis correspond to results from FCWM-base, FCWM-G, EWM, and DNS data, respectively, with DNS results shown in bold. For details of the DNS data, see \citet{Trettel2016, Yao2020}, and \citet{Gerolymos2023,Gerolymos2024a, Gerolymos2024b}.}
  \begin{ruledtabular}
  \footnotesize
  \begin{tabular}{l c c c c}
    Case & \( C_f (\times 10^{-3}) \) & \( -B_q (\times 10^{-2}) \) & \( \tilde T_c/\tilde T_w \) \\[3pt]
    \hline
    M0.74Re21092 \citep{Gerolymos2023}  & (4.890, 4.848, 5.233, \(\textbf{4.995}\)) & (1.045, 1.043, 1.087, \(\textbf{1.060}\))    & (1.096, 1.091, 1.084, \(\textbf{1.090}\)) \\
    M1.57Re25216 \citep{Gerolymos2023}  & (4.640, 4.711, 5.768, \(\textbf{4.912}\)) & (4.028, 4.085, 4.568, \(\textbf{4.140}\))    & (1.425, 1.408, 1.379, \(\textbf{1.410}\)) \\
    M0.8Re7667  \citep{Yao2020}   & (6.278, 6.287, 6.771, \(\textbf{6.290}\)) & (1.363, 1.366, 1.428, \(\textbf{1.330}\))    & (1.114, 1.111, 1.095, \(\textbf{1.109}\)) \\
    M0.8Re17000 \citep{Yao2020}   & (5.148, 5.114, 5.566, \(\textbf{5.170}\)) & (1.234, 1.233, 1.292, \(\textbf{1.210}\))    & (1.112, 1.108, 1.097, \(\textbf{1.108}\)) \\
    M0.8Re34000 \citep{Yao2020}   & (4.381, 4.383, 4.766, \(\textbf{4.470}\)) & (1.140, 1.142, 1.195, \(\textbf{1.130}\))    & (1.110, 1.104, 1.098, \(\textbf{1.109}\)) \\
    M1.5Re7667  \citep{Yao2020}\textsuperscript{\dag}   & (6.229, 6.440, 7.633, \(\textbf{6.350}\)) & (4.280, 4.358, 4.861, \(\textbf{4.260}\))    & (1.399, 1.396, 1.335, \(\textbf{1.385}\)) \\
    M1.5Re17000 \citep{Yao2020}\textsuperscript{\dag}   & (5.101, 5.199, 6.247, \(\textbf{5.300}\)) & (3.875, 3.923, 4.373, \(\textbf{3.920}\))    & (1.390, 1.382, 1.341, \(\textbf{1.388}\)) \\
    M1.5Re34000 \citep{Yao2020}   & (4.334, 4.389, 5.330, \(\textbf{4.570}\)) & (3.576, 3.624, 4.026, \(\textbf{3.660}\))    & (1.384, 1.367, 1.345, \(\textbf{1.392}\)) \\
    M1.7Re15500 \citep{Trettel2016}   & (5.210, 5.352, 6.643, \(\textbf{5.388}\)) & (4.837, 4.918, 5.590, \(\textbf{4.960}\))    & (1.497, 1.489, 1.437, \(\textbf{1.480}\)) \\
    M3.0Re24000 \citep{Trettel2016}   & (4.481, 4.868, 7.739, \(\textbf{5.048}\)) & (10.889, 11.347, 14.755, \(\textbf{11.600}\)) & (2.497, 2.505, 2.395, \(\textbf{2.491}\)) \\
    M4.0Re10000 \citep{Trettel2016}   & (5.056, 6.706, 11.960, \(\textbf{6.003}\)) & (17.169, 19.349, 28.579, \(\textbf{18.900}\)) & (3.659, 3.867, 3.474, \(\textbf{3.637}\)) \\
    M4.0Re30000                       & (3.886, 4.663, \(\textbf{--}\), \(\textbf{--}\))     & (15.026, 16.201, \(\textbf{--}\), \(\textbf{--}\))      & (3.595, 3.729, \(\textbf{--}\), \(\textbf{--}\)) \\    
  \end{tabular}
  \end{ruledtabular}
  \footnotesize
  \textsuperscript{\dag} DNS data from \citet{Yao2020} for this two cases show some discrepancies compared to the DNS from \citet{Modesti2016}. The WMLES results using FCWM-base and FCWM-G are in closer agreement with the latter.
\end{table}

\subsubsection{\label{sec:flow_filed}Flow field}
\begin{figure*}
  \includegraphics[width=\linewidth]{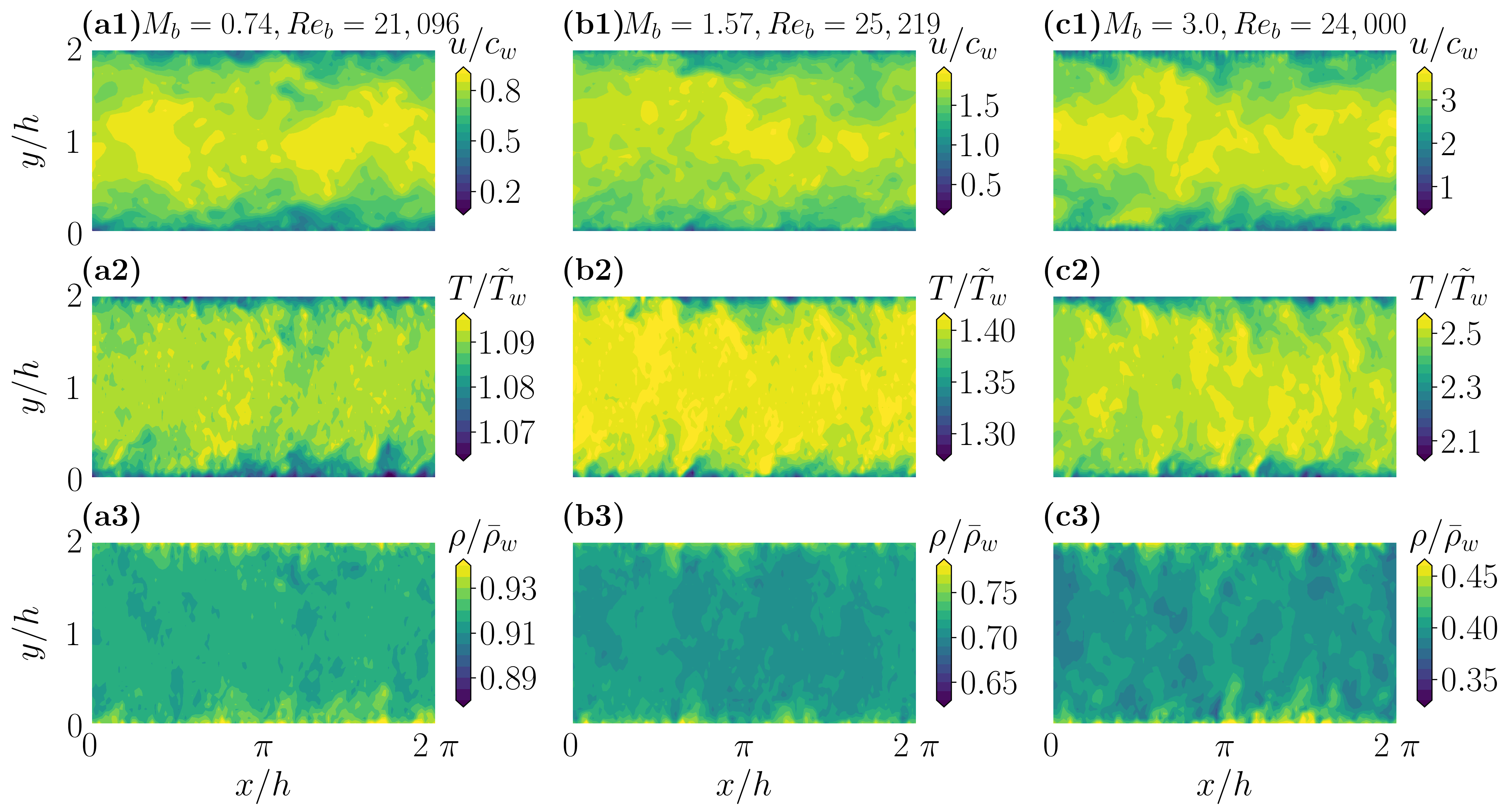}
  \caption{Contour of instantaneous velocity, temperature, and density in the X-Y plane for three conditions: \(M_b = 0.74, Re_b = 21{,}096\) (a1, a2, a3); \(M_b = 1.57, Re_b = 25{,}216\) (b1, b2, b3); and \(M_b = 3, Re_b = 24{,}000\) (c1, c2, c3), computed using FCWM-G.}
\label{fig:rhouT_contour}
\end{figure*}

Fig.~\ref{fig:rhouT_contour} presents contours of density, velocity, and temperature in the X-Y plane, computed using FCWM-G. Three representative flow conditions are shown: \(M_b = 0.74, Re_b = 21{,}092\); \(M_b = 1.57, Re_b = 25{,}216\); and \(M_b = 3.0, Re_b = 24{,}000\), corresponding to weakly, moderately, and highly compressible flows at relatively high Reynolds numbers. As shown in the figure, the contour patterns of velocity, temperature, and density are similar across all three cases, with differences primarily in magnitude. Under the isothermal wall condition, the maximum temperature is observed near the channel center, while the density reaches its maximum at the wall. As Mach number increases, this effect becomes more pronounced.

\subsubsection{\label{sec:mean_profiles}Mean profiles}
\begin{figure*}
  \includegraphics[width=\linewidth]{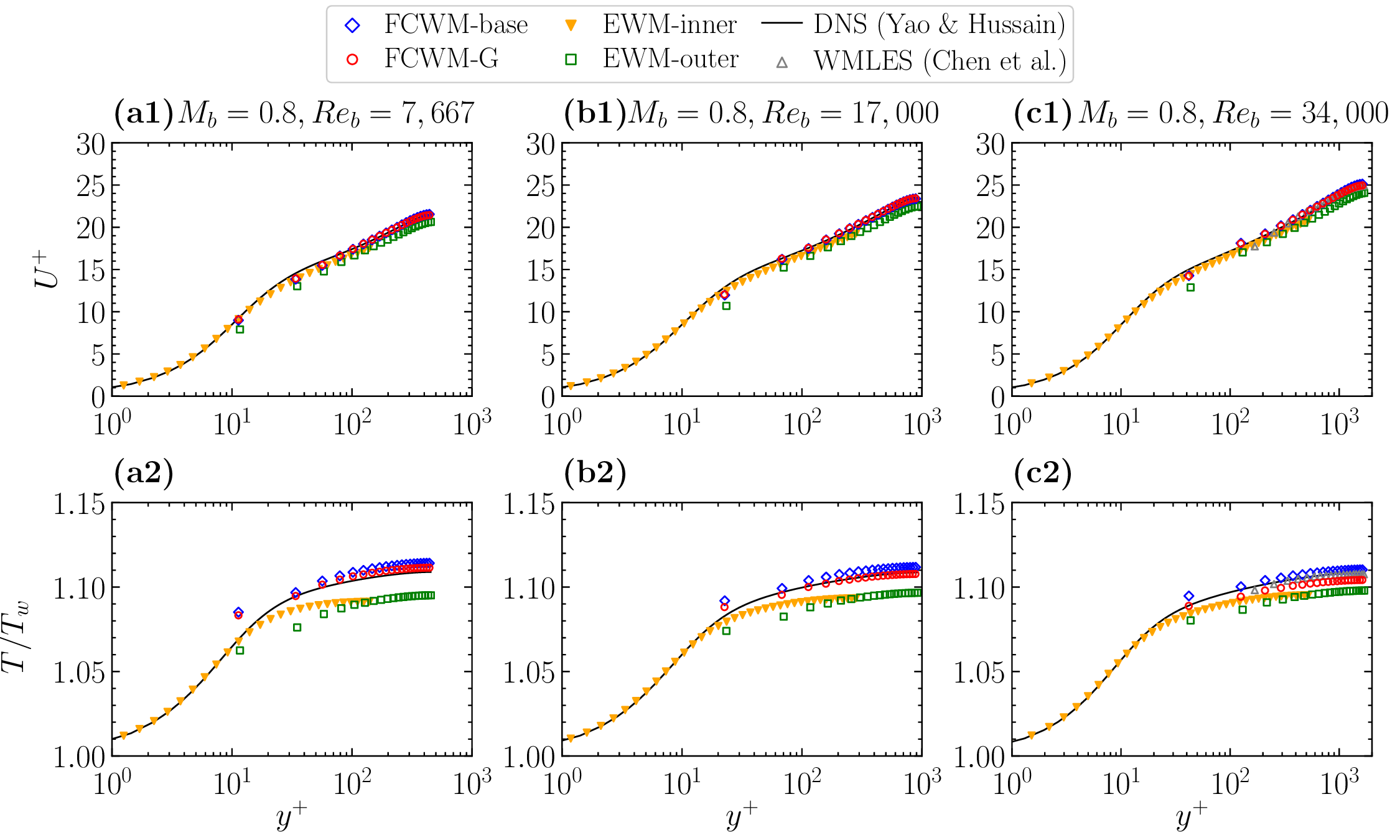}
  \caption{Velocity and temperature profiles at \(M_b = 0.8\) and various Reynolds numbers: \(Re_b = 7{,}667\) (a1, a2), \(Re_b = 17{,}000\) (b1, b2), and \(Re_b = 34{,}000\) (c1, c2). The DNS data from \citet{Yao2020} are included for comparison. In panels (c1, c2), the gray open triangles represent the outer WMLES solution by \citet{Chen2022b}, where the coupled ODEs were solved on an embedded mesh (data digitized from their published figure).}
\label{fig:YH2020_M0.8_Uplus_T_Tw}
\end{figure*}

First, we present the results for subsonic conditions. The velocity and temperature profiles for \(M_b = 0.8\) at \(Re_b = 7{,}667\), \(17{,}000\), and \(34{,}000\) are presented in Fig.~\ref{fig:YH2020_M0.8_Uplus_T_Tw}. These flow conditions correspond to those used in the DNS by \citet{Yao2020}, which serves as reference for comparison. Under this weakly compressible condition, both FCWM-base and FCWM-G yield velocity profiles in good agreement with the DNS results. In terms of temperature profile, slight discrepancies can be observed for both models, as shown in panels (a2, b2, c2). However, these discrepancies may appear amplified due to the narrow range of \(y-\)axis. Actually, we obtain \(|\epsilon_{T_c}| < 0.6\%\) for all three cases. Note that the gray open triangles in panels (c1, c2) represent the outer WMLES solution of \citet{Chen2022b}, where the coupled ODEs are solved on an embedded mesh. Our results are consistent with theirs for the case \(Re_b = 34{,}000\), even though we do not solve the ODEs. As for the EWM, the inner solutions in the viscous sublayer and buffer layer agree with the DNS results, while the outer solutions are underpredicted, especially for the temperature profiles.

\begin{figure*}
  \includegraphics[width=\linewidth]{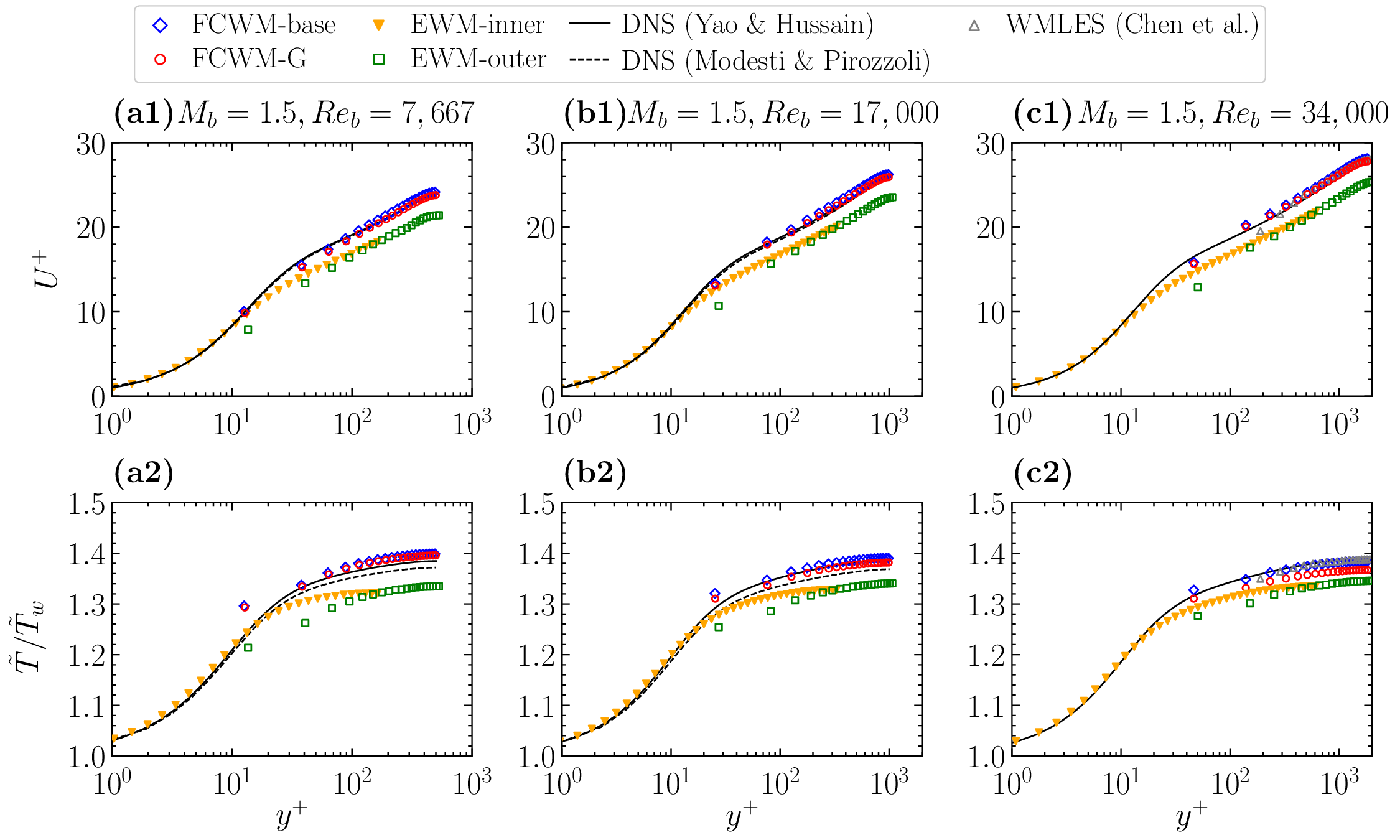}
  \caption{Velocity and temperature profiles at \(M_b = 1.5\) and various Reynolds numbers: \(Re_b = 7{,}667\) (a1, a2), \(Re_b = 17{,}000\) (b1, b2), and \(Re_b = 34{,}000\) (c1, c2). The DNS data from \citet{Yao2020} and \citet{Modesti2016} are included for comparison. In panels (c1, c2), the gray open triangles represent the outer WMLES solution by \citet{Chen2022b}, where the coupled ODEs were solved on an embedded mesh (data digitized from their published figure).}
\label{fig:YH2020_M1.5_Uplus_T_Tw}
\end{figure*}

When the Mach number increases to \(M_b = 1.5\), the computed velocity and temperature profiles using both FCWM-base and FCWM-G models agree with the DNS results of \citet{Yao2020} and \citet{Modesti2016}, as shown in Fig.~\ref{fig:YH2020_M1.5_Uplus_T_Tw}. However, it should be noted that a discrepancy exists between the two DNS datasets: the temperature profiles reported by \citet{Yao2020} are systematically higher than those of \citet{Modesti2016}, likely due to differences in grid resolution and numerical approach. For the case \(M_b = 1.5, Re_b = 34{,}000\), FCWM-G yields lower temperature than the DNS data of \citet{Yao2020}. However, when compared with the DNS results of \citet{Modesti2016} in panels (a2, b2), the temperature profile in panel (c2) is expected to have similar magnitude and closely match the FCWM-G results. In addition, our results for this case are also consistent with that of \citet{Chen2022b}. Analogous to Fig.~\ref{fig:YH2020_M0.8_Uplus_T_Tw}, the EWM provides reasonable results only in the viscous sublayer, with the outer solution for both velocity and temperature being systematically underpredicted.

\begin{figure*}
  \includegraphics[width=\linewidth]{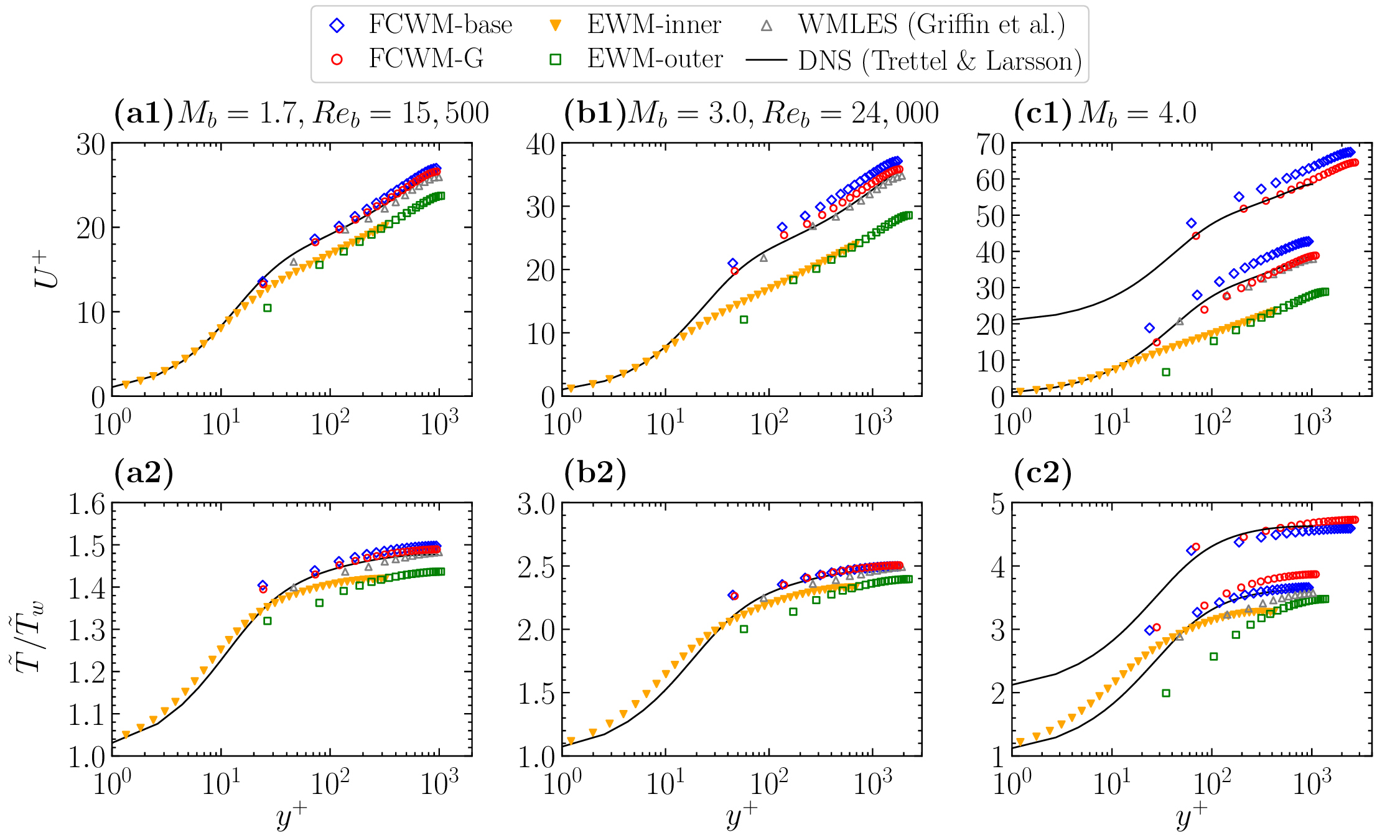}
  \caption{Velocity and temperature profile at \(M_b = 1.7, Re_b = 15{,}500\) (a1, a2); \(M_b = 3.0, Re_b = 24{,}000\) (b1, b2); and \(M_b = 4.0, Re_b = 10{,}000\) and \(M_b = 4.0, Re_b = 30{,}000\) (c1, c2). The DNS data from \citet{Trettel2016} are included for comparison. The gray open triangles denote the WMLES solution by \citet{Griffin2023} (data digitized from their published figure).}
\label{fig:TL2016_Uplus_T_Tw}
\end{figure*}

For higher Mach numbers, four cases are examined: \(M_b = 1.7, Re_b = 15{,}500\); \(M_b = 3.0, Re_b = 24{,}000\); \(M_b = 4.0, Re_b = 10{,}000\); and \(M_b = 4.0, Re_b = 30{,}000\). The first three cases are consistent with the flow conditions in the DNS of \citet{Trettel2016}. The last case is used to demonstrate the performance after eliminating low-Reynolds-number effects. Simulation results are shown in Fig.~\ref{fig:TL2016_Uplus_T_Tw}. For comparison, the wall-modeled results from \citet{Griffin2023} are also included. In their study, the incompressible momentum ODE was solved on an embedded mesh, while an inverse velocity transformation and a TV-relation were applied to compute the compressible velocity and temperature distributions.

As seen, the EWM is rather inaccurate for these high Mach number flow conditions, both velocity and temperature profiles in the outer layer presents considerable discrepancies with the DNS results. FCWM-base exhibits reduced accuracy on the velocity profiles when the Mach number reaches \(M_b = 3.0\), with performance degrading further at \(M_b = 4.0\). In contrast, FCWM-G maintains robust performance in predicting the velocity profile. However, it yields a noticeable discrepancy in the temperature profile for the case \(M_b = 4.0, Re_b = 10{,}000\). This reduced performance is likely due to the low-Reynolds-number effects \citep{Modesti2016, Griffin2023}, as the semi-local friction Reynolds number is only \(Re^*_\tau = 202\), approaching the laminar flow regime. To demonstrate this, an additional simulation is performed at \(M_b = 4.0, Re_b = 30{,}000\). Using FCWM-G, this case yields a much higher semi-local friction Reynolds number \(Re^*_\tau = 572\). The DNS results at \(M_b = 4.0, Re_b = 10{,}000\) is used for reference. As shown in panels (c1, c2), both the \(U^+\) and \(\tilde T/\tilde T_w\) profiles are significantly improved for this case. Furthermore, FCWM-G produces outer solution comparable to those of \citet{Griffin2023} without solving additional boundary layer equations. 

It is important to notice that the seemingly accurate temperature profiles from FCWM-base in the \(M_b = 3.0\) and \(M_b = 4.0\) cases are inconclusive, as they arise from potential error cancellation during the numerical integration of Eq.~(\ref{eq:Tplus_SL}) without applying the \(G^{\rho\mu}\) correction. This is evident from the relatively poor velocity predictions in panels (b1, c1).

Above results demonstrate that FCWM-base produces satisfactory velocity and temperature profiles at relatively low and moderate Mach numbers. As Mach number increases further (e.g., \(M_b \geq 3.0\)), the \(G^{\rho\mu}\) correction becomes necessary. Compared to the classical EWM, FCWM-base and FCWM-G exhibit better performance without solving the ODEs. In addition, the reference WMLES results by \citet{Chen2022b} and \citet{Griffin2023} in Figs.~\ref{fig:YH2020_M0.8_Uplus_T_Tw} to \ref{fig:TL2016_Uplus_T_Tw} are based on improved ODEs with velocity or temperature scaling corrections, which are more accurate than the conventional EWM. In the present study, the FCWM-G achieves comparable performance to these improved ODE-based wall models across a broad range of Mach and Reynolds numbers. Since the proposed FCWM does not require solving the ODEs on an embedded mesh, it is expected to significantly reduce the computational cost.

\subsubsection{\label{sec:relative_error}Relative error of key quantities}
\begin{figure*}
  \includegraphics[width=\linewidth]{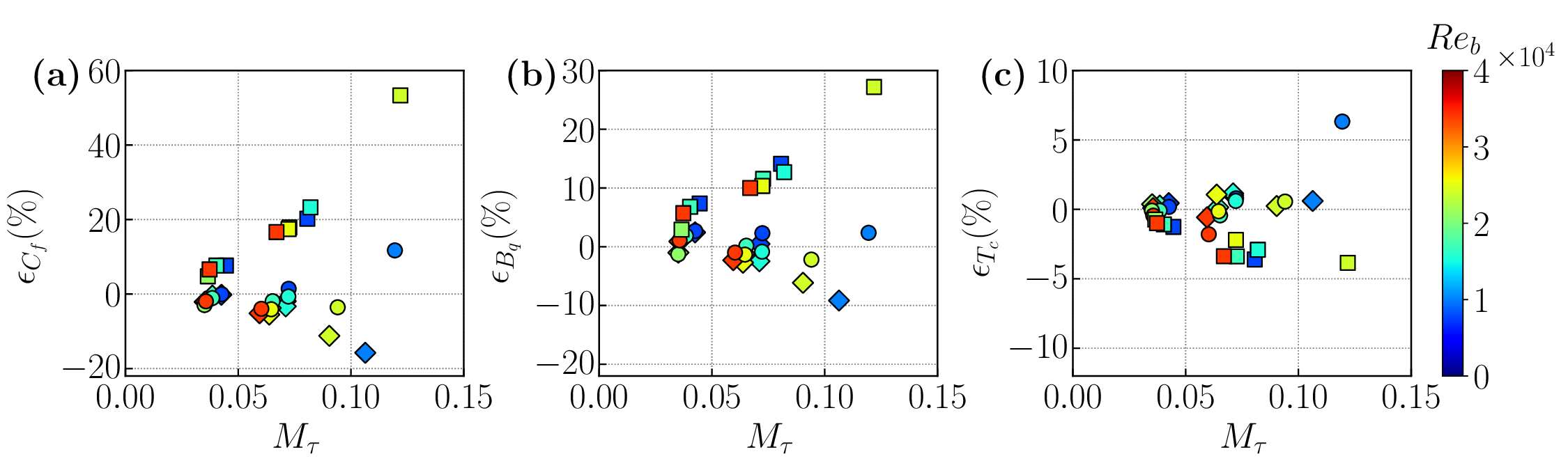}
  \caption{Relative errors in (a) the friction coefficient, (b) the non-dimensional heat flux, and (c) the centerline temperature. Symbols: filled {\(\diamond\)}—FCWM-base; filled {\large\(\circ\)}—FCWM-G; filled {\(\square\)}—EWM.}
\label{fig:error_Cf_Bq_TcTw}
\end{figure*}

The computed values of \(Re^*_\tau\), \(C_f\), \(B_q\), and \(\tilde T_c/\tilde T_w\) are listed in Table~\ref{table:cases_of_WMLES}, and the relatives errors \(\epsilon_{C_f}\), \(\epsilon_{B_q}\), and \(\epsilon_{T_c}\) are shown in Fig.~\ref{fig:error_Cf_Bq_TcTw}. The accuracy of FCWM-base, FCWM-G, and EWM is primarily influenced by \(M_\tau\), which approximately aligns with \(M_b\). Compared to the proposed wall models, EWM yields considerably larger relative errors. When \(M_\tau < 0.08\), FCWM-base performs well, but its accuracy gradually deteriorates beyond this range. In contrast, FCWM-G maintains its accuracy among the tested cases, with \(\epsilon_{C_f} < 4.1\%\), \(\epsilon_{B_q} < 2.7\%\), and \(\epsilon_{T_c} < 2.7\%\), except for \(M_b = 4.0, Re_b = 10{,}000\). As shown in Fig.~\ref{fig:TL2016_Uplus_T_Tw} (c1, c2), the reduced accuracy for this case is attributed to the low-Reynolds-number effects.

In fact, a high \(M_b\) combined with a low \(Re_b\) typically results in a low \(Re^*_\tau\) and a high \(M_\tau\). For example, the case \(M_b = 3.0, Re_b = 24{,}000\) has \(Re^*_\tau = 600\) and \(M_\tau \approx 0.097\), while the case \(M_b = 4.0, Re_b = 10{,}000\) gives \(Re^*_\tau = 203\) and \(M_\tau \approx 0.12\). This suggests that the reduced accuracy for the latter case is attributed to the low Reynolds number or large \(M_\tau\). According to \citet{Hasan2023}, \(M_\tau\) reflects compressibility effects, which in the present study are addressed through applying the revised damping function \(f^{outer}_{VD}\).

\subsubsection{\label{sec:turbulent_statistics}Turbulent statistics}
\begin{figure*}
  \includegraphics[width=\linewidth]{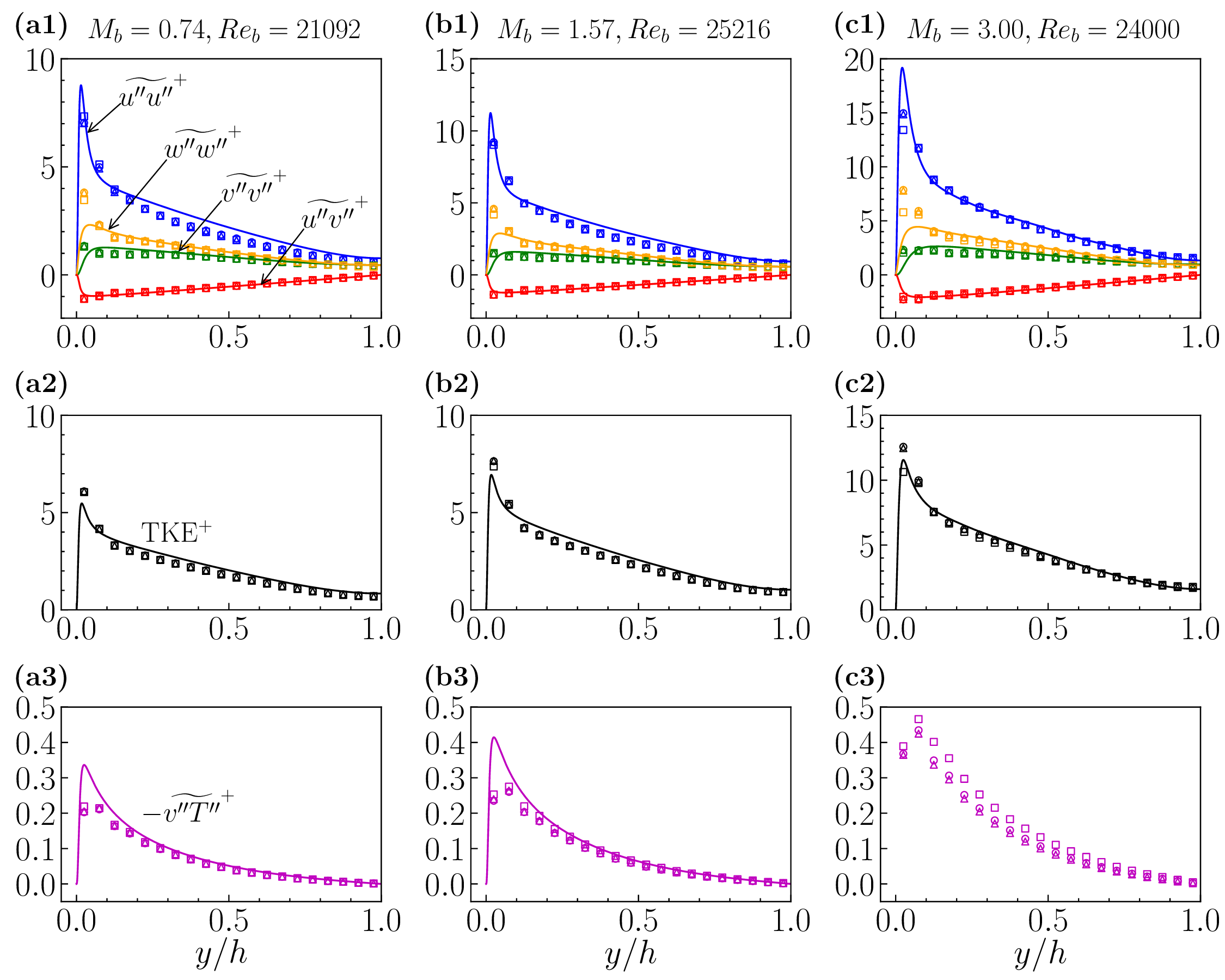}
  \caption{Turbulent statistics of WMLES at \(M_b = 0.74\), \(Re_b = 21{,}092\) (a1, a2, a3), \(M_b = 1.57\), \(Re_b = 25{,}216\) (b1, b2, b3), and \(M_b = 3.0\), \(Re_b = 24{,}000\) (c1, c2, c3). All quantities are normalized by \(u_\tau = \sqrt{\bar\tau_w/\bar\rho_w}\) and \(T_\tau = -\bar q_w/(\bar\rho_w c_p u_\tau)\). Symbols: {\(\triangle\)}—FCWM-base; {\large\(\circ\)}—FCWM-G; {\(\square\)}—EWM. Solid lines show DNS data from \citet{Gerolymos2023, Gerolymos2024a, Gerolymos2024b} and \citet{Trettel2016}. Note that the DNS data for \(\widetilde{v^{\prime\prime} T^{\prime\prime}}/(u_\tau T_\tau)\) is not available in (c3).}
\label{fig:turb_statistics}
\end{figure*}

Fig.~\ref{fig:turb_statistics} presents the turbulent statistics of the WMLES for three representative flow conditions: \(M_b = 0.74, Re_b = 21{,}092\); \(M_b = 1.57, Re_b = 25{,}216\); and \(M_b = 3.0, Re_b = 24{,}000\). The considered quantities include:
\begin{itemize}
  \item Reynolds stress, \(\widetilde{ u_i^{\prime\prime} u_j^{\prime\prime}}^+ = \widetilde{ u_i^{\prime\prime} u_j^{\prime\prime}}/u_\tau^2\);
  \item Turbulent kinetic energy, \({TKE}^+ = \frac{1}{2} \widetilde{u_i^{\prime\prime} u_i^{\prime\prime}}/u_\tau^2\);
  \item Turbulent heat flux, \(\widetilde{v^{\prime\prime} T^{\prime\prime}}^+ = \widetilde{v^{\prime\prime} T^{\prime\prime}}/(u_\tau T_\tau)\) where \(T_\tau = \bar q_w/(\bar\rho_w c_p u_\tau)\).
\end{itemize}

In contrast to the mean velocity and temperature profiles, turbulent statistics are nearly identical among FCWM-base, FCWM-G, and the conventional EWM. \citet{Catchirayer2018} reported a similar observation that both the integral and algebraic wall models produce nearly identical velocity fluctuations in subsonic and supersonic turbulent channel flows. Turbulent statistics are not resolved very well across the first few off-wall cells, which is typical for WMLES. Particularly, the \(\widetilde{ w^{\prime\prime} w^{\prime\prime}}^+\) component is overpredicted at the wall-adjacent cell. In the core region, however, the profiles agree with DNS results. According to \citet{Pope2000}, a reliable LES should resolve at least \(80\%\) of the total TKE, which is satisfied in current simulations according to panels (a2, b2, c2). Compared to the turbulent shear stress \(\widetilde{ u^{\prime\prime} v^{\prime\prime}}^+\), the turbulent heat flux \(\widetilde{v^{\prime\prime} T^{\prime\prime}}^+\) demonstrates relatively larger discrepancies near the wall. 

\subsubsection{\label{sec:computational_cost}Computational cost}
Previous results show that FCWM-base and FCWM-G yield more accurate results than the classical EWM. As the proposed FCWM does not solve the ODEs, it is of interest to compare the different wall models in term of efficiency. To this end, we compare the wall-clock times of FCWM-base, FCWM-G, and EWM.

A representative case at \(M_b = 1.57\) and \(Re_b = 25216\) to demonstrate the computational performance. For the EWM, a Newton-like approach with a relaxation coefficient of 0.5 and a tolerance of \(1\times 10^{-4}\) are applied to solve the ODEs, which typically requires \(10 \sim 15\) iterations for convergence. In contrast, the FCWM performs the entire computation in a single evaluation, with no repeated iterations. Since the convergence of the ODE solution within a single time step may slightly depend on the initial condition, the velocity and temperature fields, as well as the shear stress and heat flux are initialized using a converged WMLES realization to eliminate the influence of initial condition.

All models are implemented in JAX-Fluids \citep{Bezgin2023, Bezgin2025a}. The reported wall-clock time corresponds to the execution time of the JIT-compiled (just-in-time) version of each wall model, which runs substantially faster than standard Python code. To measure the wall-clock time, each model is first executed once to trigger JIT compilation (warm-up). Afterwards, the wall model is executed 110 times. The results from runs 11 to 110 are used to compute the minimum, maximum, mean, and standard deviation of the 100 wall-clock times. All tests are conducted on two different machines: machine~1 is equipped with an NVIDIA Quadro~K620 GPU (2048~MB GDDR5 memory, CUDA~12.4), and machine~2 with an NVIDIA RTX~A6000 GPU (49 140~MB GDDR6 memory, CUDA~12.4). The results are summarized in table~\ref{table:wall_clock_time}.

\begin{table}[t]
  \caption{\label{table:wall_clock_time} Wall-clock times of different wall models. In the EWM, the ODEs are solved on an embedded mesh with a grid of \(104 \times 40 \times 64\). In wall-normal direction, the mesh follows a hyperbolic tangent distribution with a stretching coefficient of~0.3. A Newton-like approach with a relaxation coefficient of 0.5 is applied. The iteration converges when the maximum values of both \(|\tau^{n+1}_w - \tau^{n}_w|/|\tau^{n}_w|\) and \(|q^{n+1}_w - q^{n}_w|/|q^{n}_w|\) are below \(1 \times 10^{-4}\). The speedup shown in bold is measured by the mean wall-clock time relative to the EWM as the reference.}
  \begin{ruledtabular}
  \footnotesize
  \begin{tabular}{l c c c c c c c c c c}
    \multirow{2}{*}{Wall model} & 
    \multicolumn{5}{c}{Wall-clock time [ms]: Machine 1} & 
    \multicolumn{5}{c}{Wall-clock time [ms]: Machine 2} \\
    \cmidrule(lr){2-6} \cmidrule(lr){7-11}
    & min & max & mean & std & speedup & min & max & mean & std & speedup \\[1pt]
    \hline
    EWM        & 83.098 & 84.756 & 83.197 & 0.161 & 1.00  & 6.207 & 7.005 & 6.382 & 0.204 & 1.00 \\
    FCWM-base  & 1.441  & 1.562  & 1.501  & 0.021 & \textbf{55.4}  & 0.634 & 0.785 & 0.666 & 0.031 & \textbf{9.6} \\
    FCWM-G     & 1.470  & 1.849  & 1.554  & 0.065 & \textbf{53.5}  & 0.778 & 0.926 & 0.799 & 0.016 & \textbf{8.0} \\
  \end{tabular}
  \end{ruledtabular}
\end{table}

As shown, the FCWM-base achieves an approximately 55.4\(\times\) speedup compared to the EWM on machine 1. The near-wall correction slightly increases the computational cost. However, the FCWM-G still yields a 53.5\(\times\) speedup. On machine 2, the speedups are 9.6\(\times\) and 8.0\(\times\) for FCWM-base and FCWM-G, respectively. The differences in performance between the two machines are expected and primarily result from hardware disparities. Reducing the tolerance in EWM requires more iteration steps and consequently increases the computational time. In addition, increasing the wall-normal grid resolution (\(N^{wm}_y\)), reducing the relaxation coefficient, and increasing the Mach number all lead to an increase in wall-clock times for solving the ODEs. The results in table~\ref{table:wall_clock_time} provide a representative example of the approximate comparison. In practice, the wall-clock time is further influenced by the system workload.

\section{\label{sec:discussion}Discussion}
The idea of defining near-wall modeling as a control problem was originally proposed by \citet{Nicoud2001} to overcome the numerical and modeling errors. Building on this concept, the present study extends it to compressible turbulent flows by leveraging recent developments in the compressible law of the wall. Unlike previous control-based wall models \citep{Nicoud2001,Templeton2006,Templeton2008}, the proposed FCWM employs a simpler feedback flux-control strategy that completely avoids solving the adjoint problem, thereby significantly simplifying the implementation and reducing computational complexity. This section examines the influence of mesh resolution, model parameters, the high-order term, and the compressible law of the wall. The limitations of the proposed wall model and potential improvements are also discussed.

\subsection{\label{sec:params_sensitivity}Sensitivity to mesh resolution and model parameters}
\begin{figure*}
  \includegraphics[width=\linewidth]{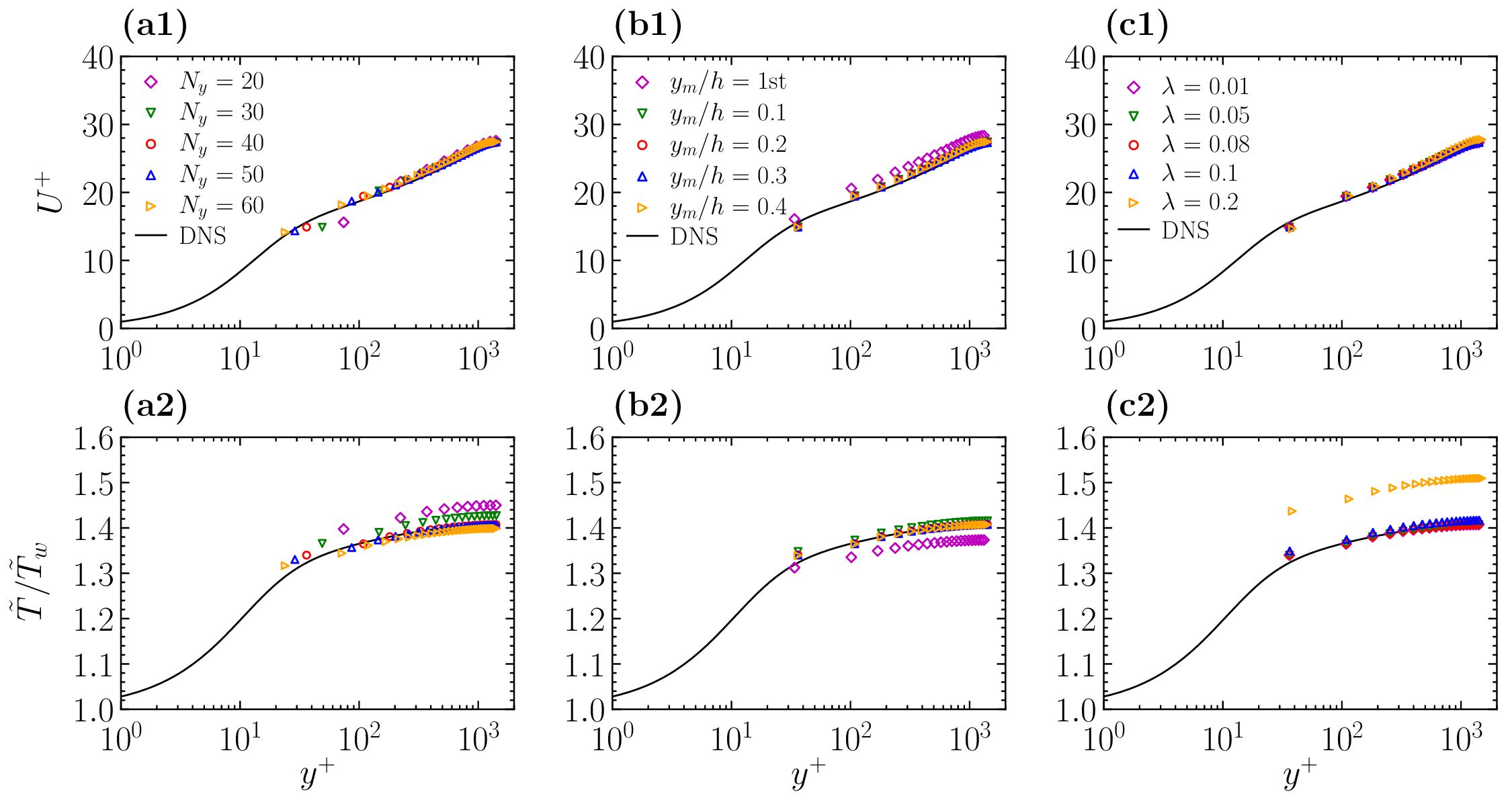}
  \caption{Sensitivity to wall-normal mesh resolution (a1, a2), matching location (b1, b2), and wall-flux relaxation coefficients \(\lambda = \lambda_\tau = \lambda_q\) (c1, c2) for the case \(M_b = 1.57, Re_b = 25{,}216\) using FCWM-G. The matching location for panels (a1,a2, c1, c2) is fixed at \(y_m/h = 0.3\). The DNS data from \citet{Gerolymos2023, Gerolymos2024a, Gerolymos2024b} are included for comparison.}
\label{fig:params_sensitivity}
\end{figure*}

Fig.~\ref{fig:params_sensitivity} presents the sensitivity of computed velocity and temperature profiles to mesh resolution, matching location \(y_m\), and wall-flux relaxation coefficients \(\lambda_\tau\) and \(\lambda_q\) for the case \(M_b = 1.57, Re_b = 25{,}216\) using FCWM-G. It is evident that the velocity distribution is less sensitive to these values than the temperature distribution.

As shown in panels (a1, a2), increasing \(N_y\) reduces the velocity discrepancy at the wall-adjacent cell center, and it also improves the overall temperature distribution. However, \(U^+\) in the outer layer is not significantly affected by \(N_y\). Beyond \(N_y \geq 40\), both profiles exhibit little variation. \citet{Larsson2016} recommend a wall-normal resolution of \(\Delta y/h \approx 0.02 \sim 0.05\), corresponding to \(N_y = 40 \sim 100\) for turbulent channel flow. In this range, the computed results show only weak dependence on the grid resolution. The choice of mesh resolution in wall-parallel directions follows the recommendation of \citet{Larsson2016}. Variations around the values applied in this study does not produce significant differences in the WMLES results; hence, they are not discussed here.

Panels (b1, b2) shows the influence of matching location using a grid resolution of \(104 \times 40 \times 64\). No significant differences are observed for \(y_m/h = 0.1 \sim 0.4\). However, choosing \(y_m\) at the first off-wall cell center results in noticeable discrepancies in both \(U^+\) and \(\tilde T/\tilde T_w\), along with relative errors of \(\epsilon_{C_f} = 14.13\%\), \(\epsilon_{B_q} = 5.64\%\), and \(\epsilon_{T_c} = 2.62\%\). The study by \citet{Kawai2012} demonstrates that the common practice of placing the matching location at the wall-adjacent cell center contributes to the well-known LLM \citep{Larsson2016, Yang2017, Maejima2024} because the LES solution there is contaminated by numerical errors. They recommend placing the matching location farther away from the wall. The same reasoning applies to the present wall model. For ease of use, \(y_m/h = 0.15 \sim 0.40\) is recommended for the proposed wall model. The first two off-wall cells should be avoided, and values of \(y_m/h > 0.5\) are also not suggested, as the coarse grid can distort the logarithmic profile (see Fig.~\ref{fig:GV2023_SL_Uplus_Tplus} (b)).

The WMLES results of using different wall-flux relaxation coefficients, \(\lambda_\tau\) and \(\lambda_q\), are presented in panels (c1, c2). In this study, the same value is applied to both parameters. The \(U^+\) profile is insensitive to variations in these parameters. However, \(\lambda = 0.2\) results in noticeable overprediction in the temperature distribution, which is reasonable as the shear stress and heat flux are not likely to fluctuate by \(20\%\) between adjacent time steps. In this study, we apply \(\lambda_\tau = \lambda_q = 0.05\).

\subsection{\label{sec:influence_of_f3}Influence of high-order term}
The temperature transformation includes high-order term, \(q^t_{tke} = -\overline{\rho v^{\prime\prime} \frac{1}{2}u^{\prime\prime}_i u^{\prime\prime}_i}\), as shown in Eq.~(\ref{eq:Tplus_SL}). According to \citet{Xu2025b}, \(q^t_{tke}\) can be neglected in mixed isothermal/adiabatic wall configuration. However, it plays a significant role in classical isothermal wall configuration. In typical WMLES, this term is only partially resolved. Consequently, two question arise: (1) How much does this under-resolution affect the logarithmic behavior of the transformed temperature profile in WMLES? (2) Can the high-order term be neglected in WMLES for the classical isothermal wall configuration?

\begin{figure*}
  \includegraphics[width=\linewidth]{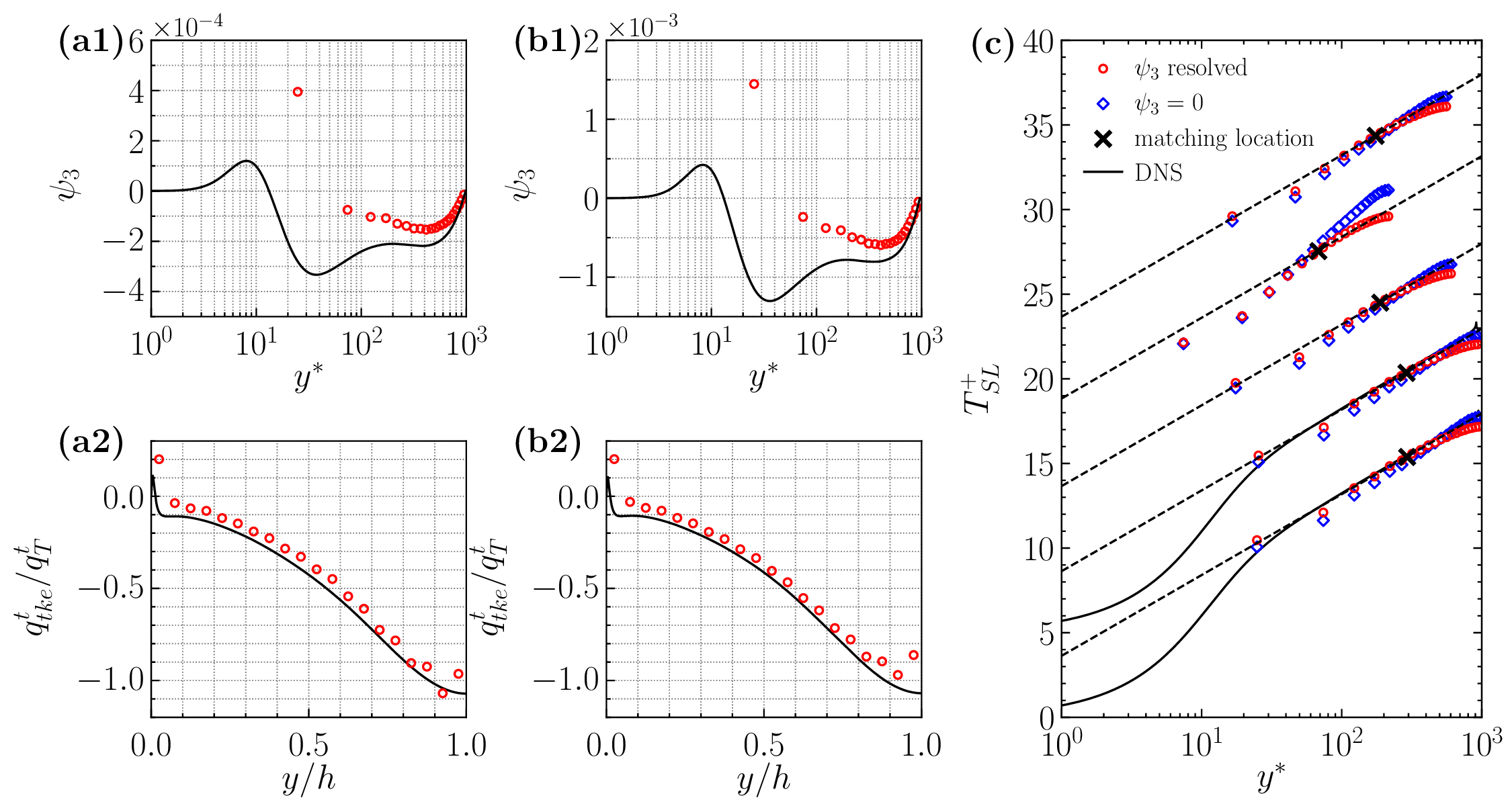}
  \caption{Influence of high-order term on the transformed temperature profile in WMLES. Panels (a1, a2) show distribution of \(\psi_3\) and \(q^t_{tke}/q^t_T\) for \(M_b = 0.74, Re_b = 21{,}092\). Panels (b1, b2) correspond to \(M_b = 1.57, Re_b = 25{,}215\). Panel (c) presents the \(T^+_{SL}\) profiles under different flow conditions, with results from bottom to top corresponding to: \(M_b = 0.74, Re_b = 21{,}092\); \(M_b = 1.57, Re_b = 25{,}215\); \(M_b = 3, Re_b = 24{,}000\); \(M_b = 4, Re_b = 10{,}000\), and \(M_b = 4, Re_b = 30{,}000\), respectively. FCWM-G is applied in all simulations. The DNS data from \citet{Gerolymos2023, Gerolymos2024a, Gerolymos2024b} are included for comparison.}
\label{fig:f3_profile}
\end{figure*}

To address the first question, Fig.~\ref{fig:f3_profile} presents the distribution of high-order term and its impact on the transformed temperature profile. DNS data for \(M_b = 0.74, Re_b = 21{,}092\) and \(M_b = 1.57, Re_b = 25{,}215\) are included for comparison. As shown in panels (a1, b1), in the near-wall region, noticeable discrepancies in \(\psi_3\) are observed between the WMLES and DNS results. In the study of \citet{Xu2025b}, the local turbulent heat conduction, defined as \(q^t_T = -\overline{\rho c_p v^{\prime\prime}T^{\prime\prime}}\), is recommended for assessing the relative importance of each component in the energy balance equation. Following this idea, the ratio \(q^t_{tke}/q^t_T\) is presented in panels (a2, b2). As shown, WMLES yields \(q^t_{tke}/q^t_T\) in close agreement with DNS data. These observations indicate that, although the WMLES cannot fully resolve \(\psi_3\), it captures the overall distribution of \(q^t_{tke}/q^t_T\). According to \citet{Xu2025b}, accounting for this ratio in the temperature transformation contributes to the formation of logarithmic profile even at relatively low Reynolds numbers. Therefore, despite the under-resolution of the high-order term in WMLES, it does not significantly affect the formation of logarithmic profile of \(T^+_{SL}\), as indicated by the red circles in panel (c).

Regarding the second question, panel (c) also includes the transformed temperature profiles computed from WMLES results by setting \(\psi_3 = 0\). Compared to using the resolved \(\psi_3\), neglecting this term leads to a slightly increased slope of \(T^+_{SL}\), as shown by the blue diamonds. However, the magnitude of \(T^+_{SL}\) near the matching location does not exhibit significant difference at high Reynolds numbers. The discrepancy observed in the case \(M_b = 4.0, Re_b = 10{,}000\) is likely due to low-Reynolds-number effects. These observations suggest that, at sufficiently high Reynolds numbers, neglecting the high-order term does not significantly affect the \(T^+_{SL}\) profile, even at a Mach number as high as \(M_b = 4.0\).

\begin{figure*}
  \includegraphics[width=\linewidth]{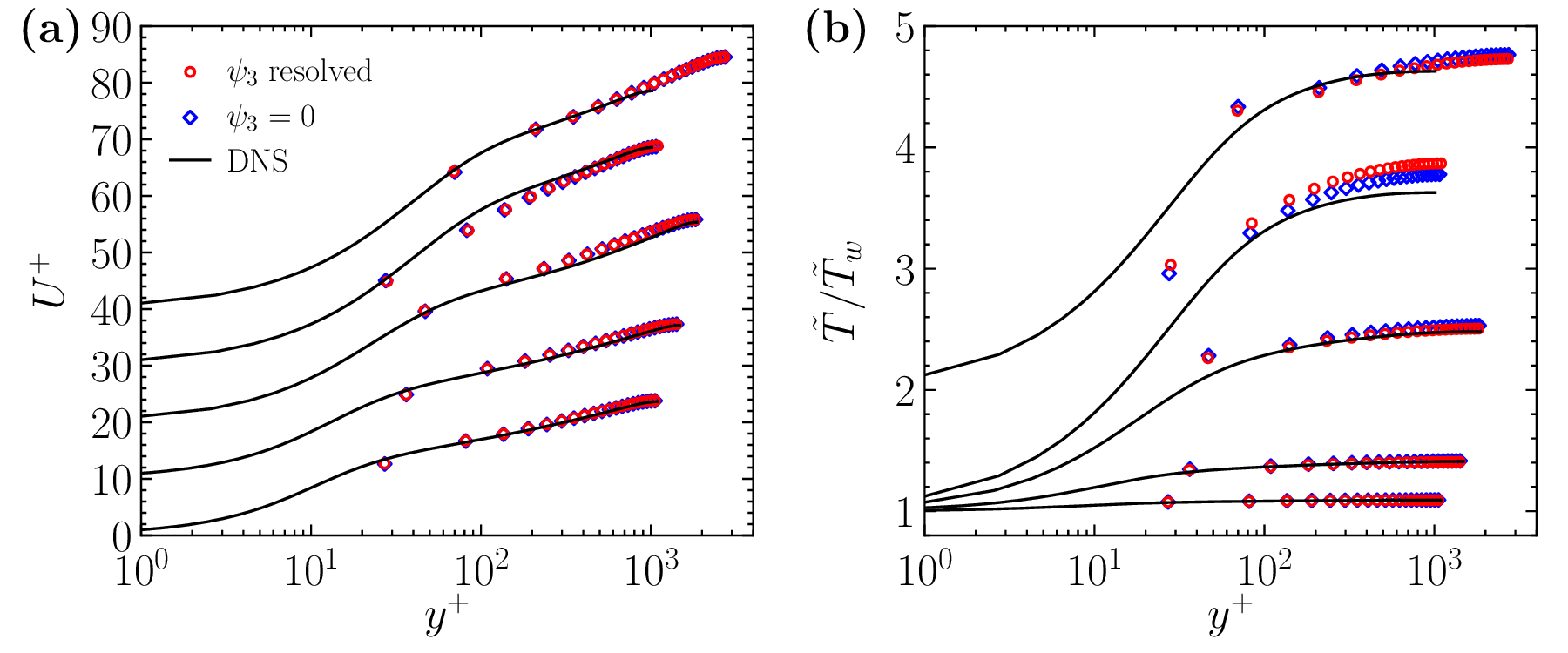}
  \caption{Influence of neglecting the high-order term on WMLES results. (a) \(U^+\) profiles, (b) \(\tilde T/\tilde T_w\) profiles. Results from bottom to top correspond to: \(M_b = 0.74, Re_b = 21{,}092\); \(M_b = 1.57, Re_b = 25{,}215\); \(M_b = 3, Re_b = 24{,}000\); \(M_b = 4, Re_b = 10{,}000\); and \(M_b = 4, Re_b = 30{,}000\), respectively. In panel (a), the results are shifted upward by multiple of 10 units for clarity. In panel (b), the results for \(M_b = 4, Re_b = 30{,}000\) are shifted upward by 1 unit. The DNS data for \(M_b = 4, Re_b = 10{,}000\) are also used as a reference for the \(M_b = 4, Re_b = 30{,}000\) case. DNS data from \citet{Trettel2016} and \citet{Gerolymos2023, Gerolymos2024a, Gerolymos2024b} are included for comparison.}
\label{fig:f1f2RhoMu_Uplus_T_Tw}
\end{figure*}

To further investigate the influence of neglecting \(\psi_3\) on the WMLES results, simulations with \(\psi_3 = 0\) are performed for the same cases shown in Fig.~\ref{fig:f3_profile} (c). The results are presented in Fig.~\ref{fig:f1f2RhoMu_Uplus_T_Tw}. For comparison, results from DNS and WMLES using the resolved \(\psi_3\) are also included. Consistent with the findings in Fig.~\ref{fig:f3_profile}, at high Reynolds numbers, neglecting the high-order term does not lead to significant difference in either the velocity or temperature profiles. The relative discrepancies in the computed \(C_f\), \(B_q\), and \(\tilde T_c/\tilde T_w\) are no more than \(1\%\), except for the case \(M_b = 4.0, Re_b = 10{,}000\) where slightly larger deviations occur due to low-Reynolds-number effects. These deviations reduce at higher Reynolds numbers, as demonstrated by the case at \(M_b = 4.0, Re_b = 30{,}000\).

Finally, it should be noted that the study of \citet{Xu2025b} highlights the importance of including the high-order term in the temperature transformation for the configuration used here. This conclusion does not conflict with the above results, as most available DNS data employed in \citet{Xu2025b} correspond to low Reynolds numbers. Since WMLES is typically intended for high-Reynolds-number flows, under-resolving or completely neglecting \(\psi_3\) have only limited impact on the WMLES results. Nevertheless, we recommend including \(\psi_3\) in the wall model to allow for a broader range of logarithmic profiles, thereby enhancing the robustness of the wall model.

\subsection{\label{sec:challenges_from_compressible_LoW}Challenges from the compressible law of the wall}
Both the approach of \citet{Nicoud2001} and the present wall model rely on the law of the wall. In principle, WMLES is intended for high-Reynolds-number flows, where the log-law asymptotically approach a "universal" form characterized by a constant slope and intercept. However, for compressible turbulent channel flow, reliable data at sufficiently high Reynolds numbers remain limited. The DNS datasets by \citet{Yao2020}, \citet{Lusher2022}, and \citet{Gerolymos2023, Gerolymos2024a, Gerolymos2024b} are among the few publicly available resources with \(Re^*_\tau\) reaching or exceeding 1000, where a clear logarithmic region can be observed. Therefore, a practical challenge in the compressible WMLES is to account for the low-Reynolds-number effects. In the present study, these effects are directly connected to the variation in the log-law intercepts, \(B\) and \(B_T\). This issue is addressed by employing the fitted relations provided in Eq.~(\ref{eq:loglaw_B_BT}). The results in Sec.~{\ref{sec:application}} demonstrate the effectiveness of this approach.

To assess the impact of uncertainties in the fitted formulas, we conducted a sensitivity study by manually perturbing the fitted \(B\) and \(B_T\) values in Eq.~(\ref{eq:loglaw_B_BT}) using shifts \(\delta B \in [-0.13, 0, 0.13], \delta B_T \in [-0.1, 0, 0.1]\), which align with the \(\pm \, rms\) error margin shown in Fig.~\ref{fig:loglaw_B_BT}. This results in \(3 \times 3\) combinations of \((\delta B, \delta B_T)\), with \((\delta B, \delta B_T) = (0, 0)\) corresponding to the baseline. Simulation results using FCWM-G for the case \(M_b = 1.57, Re_b = 25{,}216\) show only minor discrepancies. Compared to the baseline, the maximum relative deviations in the computed \(C_f\), \(B_q\), and \(\tilde T_c/\tilde T_w\) across the eight perturbed cases are \(1.6\%\), \(0.6\%\), and \(0.5\%\), respectively. These results suggest that the uncertainties in the fitted formula for \(B\) and \(B_T\) have limited impact on the WMLES outcomes.

The present control-based approach is modular, making it feasible to accommodate other velocity and temperature transformations \citep{Patel2016, Volpiani2020, Griffin2021, Patel2017, Chen2022a, Huang2023, Cheng2024a}. Furthermore, with the continuous growth of high resolution datasets and ongoing research, more advanced formulations of the compressible law of the wall are expected to be developed in the future, which can further enhance the performance of the present approach.

\subsection{\label{sec:more_advanced_control_strategy}Limitations and potential improvements}
In principle, the velocity and temperature transformations in Eqs.~(\ref{eq:Uplus_SL}) and (\ref{eq:Tplus_SL}) require high resolution data, which conflicts with the inherently coarse near-wall resolution of WMLES. Based on the tested cases in this study, when \(M_b \leq 1.7\), this under-resolution does not lead to significant errors in the baseline wall model, FCWM-base. For higher Mach numbers, the \(G^{\rho\mu}\) correction is required to account for the drastic variations in fluid properties within the viscous sublayer and buffer layer. The resulting wall model, FCWM-G, demonstrates good performance across a wide range of Mach and Reynolds numbers. Nonetheless, there are still observable discrepancies in the velocity and temperature profiles, as well as in the computed \(C_f\), \(B_q\), and \(\tilde T_c/\tilde T_w\) when compared to DNS data. To further enhance the performance of the proposed wall model, the following aspects can be considered.

First, more physical insights can be incorporated into the "loss function". The present flux-control strategy relies solely on \(\Delta U^+_{SL}\) and \(\Delta T^+_{SL}\) at the matching location, as indicated in Eq.~(\ref{eq:delta_Uplus_Tplus}). Other physically important information from the outer LES solution is not effectively utilized, such as Reynolds stress and profiles above and below the matching location. Importantly, both velocity and temperature transformations are inherently coupled: the velocity distribution directly influences the \(T^+_{SL}\) profile through \(\psi_2\) and \(u^+\) in the denominator of Eq.~(\ref{eq:Tplus_SL}). In turn, the temperature transformation affects the velocity transformation indirectly by influencing the temperature field, which determines the profiles of density and viscosity that enter the velocity transformation. However, in order to simplify the implementation and reduce computational cost, both velocity and temperature transformations are implemented separately to determine the mean shear stress and heat flux in the present wall model. As a result, it cannot adequately account for these coupling between velocity and temperature transformations. 

Second, a more advanced approach can be explored for prescribing the local shear stress and heat flux. The shifted boundary condition in Eq.~(\ref{eq:shifted_boundary_condition}) assumes a close cross-correlation between the wall shear stress (heat flux) and the velocity (temperature) at the first off-wall cell center. Compared to DNS and WRLES results, this treatment results in larger cross-correlation between the wall and the corresponding \(y-\)plane. In contrast, the methods of \citet{Nicoud2001} and \citet{Bae2022} directly adjust the local shear stress rather than the mean value, which could also be incorporated into the present wall model.

Third, the coupling effects between the wall model and outer LES solver can be accounted. In addition to the wall model itself, results of WMLES also depend on the outer LES solver. As indicated in Fig.~\ref{fig:GV2023_Uplus_T_Tw_fVD}, using a revised damping function \(f^{outer}_{VD}\) in ALDM helps reduce the velocity discrepancy at the wall-adjacent cell center and improve the temperature prediction. A recent study by \citet{Liu2025} reveals that correcting the SGS viscosity in the near-wall region using wall shear stress from the wall model effectively reduces the LLM. In ALDM, adjusting the damping function has a similar effect to modifying the SGS viscosity in the near-wall region. Therefore, it would be valuable to take into consideration the coupling between the proposed wall model and the outer LES solver.

Addressing the above aspects will inevitably increase the implementation complexity and computational cost of the flux-control strategy. In this regard, differentiable CFD solvers like JAX-Fluids \citep{Bezgin2023, Bezgin2025a}, which leverage automatic differentiation \citep{Baydin2018} and enable end-to-end optimization \citep{Bezgin2025b}, offer promising opportunities to develop more advanced flux-control strategies. 

\subsection{\label{sec:more_general_configuration}Application to more general configuration}
As introduced earlier, the proposed FCWM consists of three components: (1) the compressible law of the wall, (2) a feedback flux-control strategy, and (3) a shifted boundary condition \citep{Piomelli1989}. The present study only evaluates the performance in turbulent channel flow, as the temperature transformation given in Eq.~\eqref{eq:Tplus_SL} is specifically formulated for this configuration. For more general flow configurations, such as flat plate turbulent boundary layer with or without pressure gradient, a well-established temperature transformation is still lacking, which is the primary challenge in evaluating the proposed FCWM for these configurations. Nevertheless, both the feedback flux-control strategy and the shifted boundary condition are general, and can be readily applied to other wall-bounded turbulent flows. The potential extension of the temperature transformation in Eq.~\eqref{eq:Tplus_SL} to more general flows may be achieved through parameters \(\psi_1\), \(\psi_2\), and \(\psi_3\) in Eq.~\eqref{eq:beta_f1f2f3}. Detailed discussions can be found in \citet{Xu2025b}. Once a more advanced temperature transformation is available, the FCWM framework can be readily applied to these flow configurations.

\section{\label{sec:conclusion}Conclusion}
In this study, a flux-controlled wall model for LES of wall-bounded turbulent flow is proposed. It leverages the velocity and temperature transformations and employs a simplified feedback flux-control strategy to adjust the wall shear stress and heat flux. To account for the sharp variation of fluid properties in the near-wall region, the \(G^{\rho\mu}\) correction is introduced. Two versions of the wall model are proposed: FCWM-base and FCWM-G. Both models are evaluated via WMLES of turbulent channel flow across a wide range of Mach and Reynolds numbers, including quasi-incompressible cases with \(M_b = 0.1\) and \(Re_\tau = 180 \sim 10{,}000\), and compressible cases with \(M_b = 0.74 \sim 4.0\) and \(Re_b = 7667 \sim 34{,}000\). The simulation results show good agreement with DNS data in the mean velocity and temperature profiles. For \(M_b \leq 1.7\), FCWM-base performs well, while the \(G^{\rho\mu}\) correction becomes necessary at higher Mach numbers. Across the tested cases, FCWM-G achieves \(|\epsilon_{C_f}| < 4.1\%\), \(|\epsilon_{B_q}| < 2.7\%\), and \(|\epsilon_{T_c}| < 2.7\%\) for \(M_b \leq 3\) when compared with DNS results. The slightly reduced accuracy observed at \(M_b = 4, Re_b = 10{,}000\) is likely due to low-Reynolds-number effects, which improves at higher Reynolds numbers. FCWM-base and FCWM-G produce no significant differences in Reynolds stress and turbulent heat flux. The FCWM demonstrates similar accuracy to the improved ODE-based equilibrium wall models by \citet{Chen2022b} and \citet{Griffin2023} in compressible turbulent channel flows. Compared to the conventional ODE-based equilibrium wall model, the proposed FCWM achieves higher accuracy without solving the boundary layer equations, thereby reducing computational cost. For proper implementation of the wall model, a matching location of \(y_m/h = 0.15 \sim 0.4\) and wall-flux relaxation coefficients of \(\lambda_\tau = \lambda_q = 0.01 \sim 0.08\) are recommended. Although WMLES cannot fully resolve the high-order term in \(\psi_3\), it captures the overall distribution of \(q^t_{tke}/q^t_T\), which contributes to the formation of the logarithmic profile of \(T^+_{SL}\) and supports the flux-control strategy. At high Reynolds numbers, the high-order term does not significantly affect the distribution of \(T^+_{SL}\), and completely neglecting it (\(\psi_3 = 0\)) makes no substantial difference in the WMLES results. Nevertheless, we recommend including \(\psi_3\) in the model to allow for a broader range of logarithmic profiles, thereby enhancing the robustness of the wall model.

The modular structure of the control-based approach readily accommodates alternative velocity and temperature transformations. To further enhance the performance, additional physical insights of the flow can be incorporated into the flux-control strategy. Although the present work focuses on turbulent channel flow, the proposed flux-control strategy can be extended to more general wall-bounded turbulent flows. The primary challenge lie in developing a compressible temperature transformation for such cases, which will be addressed in future investigations.

\begin{acknowledgments}
The first author gratefully acknowledges financial support from the China Scholarship Council (No.202006320042). The authors sincerely acknowledge the members of the JAX-Fluids group in the Chair of Aerodynamics and Fluid Mechanics of TUM for their valuable discussions and kind help.
\end{acknowledgments}

\section*{AUTHOR DECLARATIONS}
\subsection*{Conflict of Interest}
The authors have no conflicts to disclose.

\subsection*{Author Contributions}
\textbf{Youjie Xu}: Conceptualization (lead); Data curation (lead); Formal analysis (lead); Investigation (lead); Methodology (lead); Software (lead); Validation (lead); Visualization (lead); Writing-original draft (lead). \textbf{Steffen J. Schmidt}: Funding acquisition (equal); Project administration (equal); Resources (equal); Supervision (equal); Writing-review \& editing (equal). \textbf{Nikolaus A. Adams}: Funding acquisition (equal); Project administration (equal); Resources (equal); Supervision (equal); Writing-review \& editing (equal).

\section*{Data Availability Statement}
The DNS data that used as reference in this study are available in the cited literature. The WMLES data can be obtained from the corresponding author upon reasonable request.

\section*{Author ORCID.}
Youjie Xu \href{https://orcid.org/0009-0006-8445-3200}{https://orcid.org/0009-0006-8445-3200};

Steffen J. Schmidt \href{https://orcid.org/0000-0001-6661-4505}{https://orcid.org/0000-0001-6661-4505};

Nikolaus A. Adams \href{https://orcid.org/0000-0001-5048-8639}{https://orcid.org/0000-0001-5048-8639}.

\appendix*
\section{Implementation of EWM}\label{app:ewm}
Within the EWM framework, the simplified ODEs for momentum and energy balances are given by \citep{Kawai2012, Larsson2016}:
\begin{equation}\label{eq:EWM_U}
  \frac{d}{dy} \left[ \left( \mu + \mu_{t,\mathrm{wm}} \right) \frac{dU}{dy} \right] = 0,
\end{equation}
\begin{equation}\label{eq:EWM_T}
  \,\frac{d}{dy} \left[ \left( \mu + \mu_{t,\mathrm{wm}} \right) U \frac{dU}{dy} + 
  c_p \left( \frac{\mu}{\mathrm{Pr}} + \frac{\mu_{t,\mathrm{wm}}}{\mathrm{Pr}_{t,\mathrm{wm}}} \right)\frac{dT}{dy} \right]
  = 0.
\end{equation}

Here \(U\) denotes the mean wall-parallel velocity, and \(T\) represents the mean temperature. \(\kappa = 0.41\) and \(\bar\mu/\bar\mu_w = (\tilde T/\tilde T_w)^{0.7}\) are applied as in the FCWM. In the EWM, the molecular and turbulent Prandtl number are set to be \(Pr = 0.7\) and \(Pr_{t,wm} = 0.9\), respectively, consistent with previous studies \citep{Larsson2016, Griffin2023}. The wall model eddy viscosity is given by \(\mu_{t,wm} = \kappa \rho \sqrt{\tau_w/\rho} \, y \, D\), with the damping function given by \(D = \left[ 1 - \exp\!\left(-y^+/17\right) \right]^2\).

To solve the ODEs, a one-dimensional stretched mesh is applied in the wall model. The mesh distribution follows the hyperbolic tangent function, as given below:
\begin{equation}\label{eq:hyperbolic_tanh_mesh}
y = \frac{L}{2} \left[ 1 - \frac{\tanh\!\big(\beta(1 - 2\xi)\big)}{\tanh(\beta)} \right].
\end{equation}

Here, \(L = 2 \, y_m\), \(\beta = 3.0\), \(\xi \in [0, 1]\). The wall-adjacent mesh satisfies \(\Delta y^+_w < 1\). Following \citet{Griffin2023}, a Newton-like iterative method is used to solve the ODE. The iteration converges when the maximum values of both \(\epsilon_\tau = |\tau^{n+1}_w - \tau^{n}_w| / |\tau^{n+1}_w| < 1 \times 10^{-4}\) and \(\epsilon_q = |q^{n+1}_w - q^{n}_w| / |q^{n+1}_w| < 1 \times 10^{-4}\) are satisfied or after reaching the maximum iteration limit.

\begin{figure*}[ht]
  \includegraphics[width=0.9\linewidth]{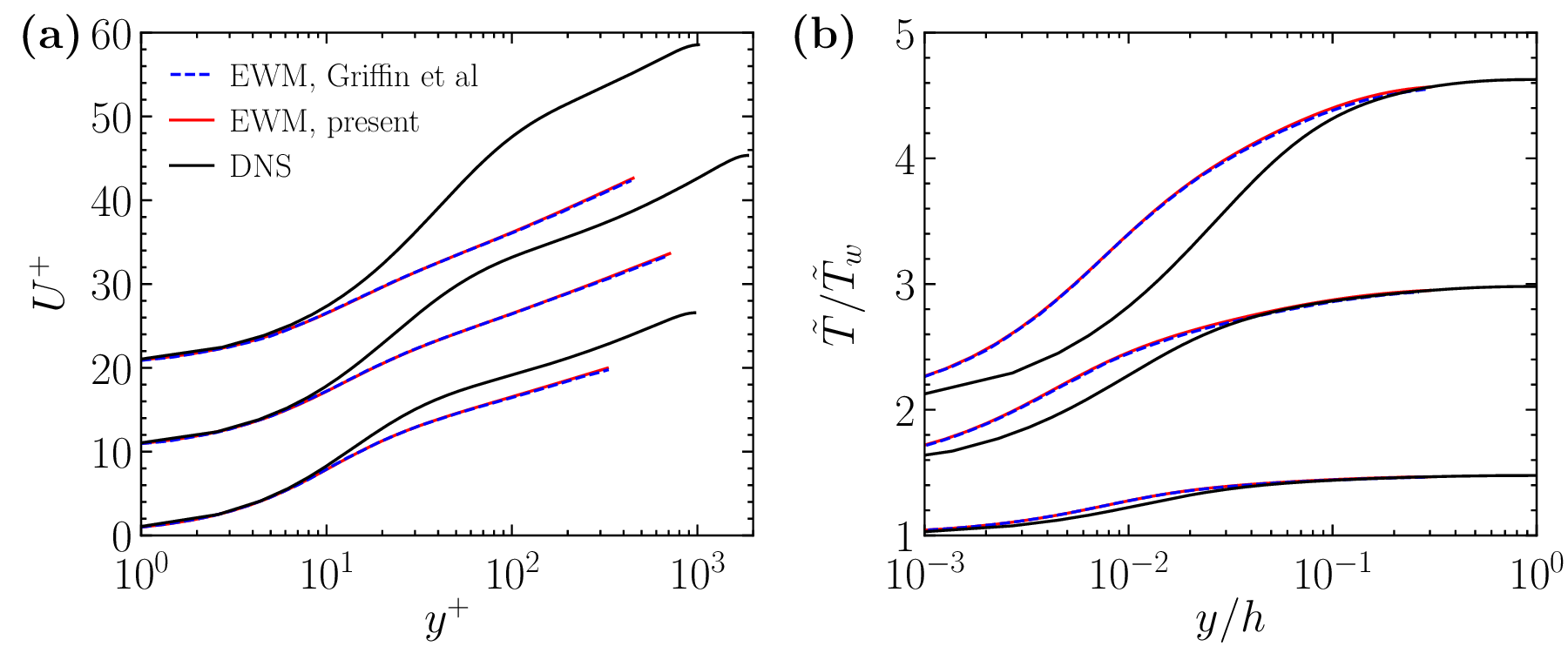}
  \caption{Comparison of \emph{a priori} test. The curves are shifted upward in multiple of 10 units in (a) and 1 unit in (b). The three cases are \(M_b = 1.7, Re_b = 15500\), \(M_b = 3.0, Re_b = 24000\), and \(M_b = 4.0, Re_b = 10000\). The DNS data from \citet{Trettel2016} are included for comparison. The \emph{a priori} test results of \citet{Griffin2023} are included for comparison.}
\label{fig:validate_EWM}
\end{figure*}

To verify the correct implementation of EWM, we conduct the \emph{a priori} test and compare our results with those of \citet{Griffin2023} across the same cases. In this test, the velocity and temperature at the matching location are directly extracted from the DNS data of \citet{Trettel2016}, hence eliminating the influence of LES solver. The matching location is chosen as in \citet{Griffin2023}, with \(y_m/h = 0.3\). Note that the dynamic viscosity satisfies \(\bar\mu/\bar\mu_w = (\tilde T/\tilde T_w)^{0.75}\) in the DNS of \citet{Trettel2016}. The \emph{a priori} test results are presented in Fig.~\ref{fig:validate_EWM}. For comparison, the \emph{a priori} test results of \citet{Griffin2023} for the same cases are also included. As shown, our results are in close agreement with those of \citet{Griffin2023}, therefore verifying the correct implementation.

Note that in the FCWM, we apply \(Pr_t = 0.85\) for the temperature log-law in the overlap region, which is slightly different from the value employed in the EWM. The power index for the dynamic viscosity is set to 0.7 in Sections \ref{sec:methodology}, \ref{sec:improved_wall_model}, and \ref{sec:application} for consistency throughout the study. This choice generally does not lead to significant differences in the simulation results within the considered flow conditions.

\bibliography{reference}

\end{document}